\documentclass[10pt,onecolumn,twoside]{IEEEtran}

\usepackage{generic}
\usepackage{amsmath,amssymb,amsfonts,soul}
\DeclareMathOperator{\Tr}{Tr}
\usepackage{graphicx}
\usepackage{textcomp}

\def\BibTeX{{\rm B\kern-.05em{\sc i\kern-.025em b}\kern-.08em
    T\kern-.1667em\lower.7ex\hbox{E}\kern-.125emX}}
    
\usepackage{amsthm}
\usepackage{graphics} 
\usepackage{epsfig} 
\usepackage{times} 

\usepackage[maxbibnames=99,giveninits=true, sorting=none]{biblatex}
\usepackage{bm}
\usepackage{xcolor}
\usepackage{booktabs}
\usepackage{siunitx}
\usepackage{hyperref}

\usepackage{caption}
\usepackage{lipsum}
\usepackage{multirow}
\usepackage{diagbox} 
\usepackage{subcaption}
\renewcommand*{\thefootnote}{\fnsymbol{footnote}} 

\newcommand{\R}{\mathbb{R}}
\newcommand{\re}{\text{Re}}

\newcommand{\Pro}{\mathbb{P}}
\newcommand{\io}{\iota_{\varepsilon}}
\newcommand{\cass}{\mathcal{A}^{i,j}_{\varepsilon}}
\newcommand{\casm}{\mathcal{A}^{\mathcal{I}_m,j}_{\varepsilon}}
\newcommand{\srisk}{\mathcal{A}^{j}_{\varepsilon}}
\newcommand{\sigi}{\sigma_i}
\newcommand{\sigj}{\sigma_j}

\usepackage[T1]{fontenc}
\newcommand{\textincon}[1]{%
{\fontfamily{zi4}\selectfont #1}}
\newcommand{\VAR}{{\textrm{\large \textincon{V$@$R}}}}
\newcommand{\AVAR}{{\textrm{\large \textincon{AV$@$R}}}}

\newtheorem{theorem}{\bf Theorem}
\newtheorem{lemma}{\bf Lemma}
\newtheorem{corollary}{\bf Corollary}
\newtheorem{remark}{\bf Remark}

\title{\huge \bf
Risk of Cascading Collisions in Network of Vehicles \\  with Delayed Communication
}

\author{Guangyi Liu, Christoforos Somarakis, and Nader Motee$^{1}$

\thanks{G. Liu and N. Motee are with the Department of Mechanical Engineering and Mechanics, Lehigh University, Bethlehem, PA, 18015, USA. {\tt\small \{gliu,motee\}@lehigh.edu}.
C. Somarakis is Senior Scientist with the Applied Mathematics Group, Merck \& Co {\tt\small christoforos.somarakis@merck.com}.
}
}

\begin{document}

\maketitle

\begin{abstract}
This paper establishes and explores a framework to analyze the risk of cascading failures in a platoon of autonomous vehicles, accounting for communication time-delays and input uncertainty. Our proposed framework yields closed-form expressions for cascading collisions, which we quantify using the coherent Average Value-at-Risk ($\AVAR$) to assess the cascading effect of vehicle collisions within the platoon. We investigate how factors such as network connectivity, system dynamics, communication delays, and uncertainty contribute to the emergence of cascading failures. Our findings are extended to standard communication graphs with symmetries, allowing us to evaluate the risk of cascading collisions from a platoon design perspective.  Furthermore, by discovering the boundedness of the inter-vehicle distances, we reveal the best achievable risk of cascading collision with general graph topologies, which is further specified for special communication graph, such as the complete graph. Our theoretical results pave the way for the development of a safety-aware framework aimed at mitigating the risk of cascading collisions in vehicle platoons.
\end{abstract}


\section{Introduction}

Uncertainties and time delays are intrinsic elements that persistently manifest in networked control systems. The uncertainties arise from various factors, including sensor inaccuracies, environmental disturbances, and limitations in modeling the dynamic behavior of the controlled processes. Meanwhile, time delays emerge from multiple sources, such as communication latencies, processing times, and transmission constraints. The uncertainties and time delays in networked control systems introduce significant challenges for achieving controllability and ensuring safety. The unpredictability of system behavior due to uncertainties limits the efficacy of control strategies, potentially leading to undesired states or unstable behavior \cite{zhou1996robust}. Additionally, the time delays can disrupt the closed-loop control and compromise system stability \cite{gu2003stability}. There are numerous techniques available to measure the performance of a networked control system and assess the risk of failures in {  the} system network. Some notable examples include studies in financial networks \cite{acemoglu2015systemic}, large-scale networks \cite{bamieh2012coherence}, vehicular networks \cite{grunberg2017determining}, power networks \cite{dorfler2012synchronization}, and general networked control systems \cite{sandberg2015cyberphysical, sandberg2022secure, siami2017growing, amini2022space}.

{ In this paper, we explore an alternative perspective. Considering the stochastic nature of real-world platoons, our working hypothesis is that failures and undesirable events are inevitable. In particular, in high-dimensional network controlled systems, even slight local failures can propagate through the system's interaction mechanisms, leading to global failure. Therefore, our focus is on studying cascading failures \cite{7438924, zhang2018cascading, zhang2019robustness,liu2022risk,liu2023cascading} in the networked control system of autonomous vehicle platooning. }

{ Hence, it is essential to investigate tools for networked control systems that can evaluate the networks' ability to prevent the spread of cascading failures. To accomplish this, we develop a theoretical framework based on systemic risk analysis \cite{Somarakis2020b}, with the aim of assessing the risk of cascading failures. Our goal is to highlight how systemic failures in specific regions of the network can trigger additional failures across the network. By quantifying the impact of such failures, we can gain valuable insights into system design and recognize the vulnerabilities of networked systems to cascading failures. Inspired from the extensive analysis conducted in various networked control systems using the risk measures \cite{Somarakis2020b, Somarakis2019g}, we adopt both $\VAR$ \cite{rockafellar2000optimization} and Average Value-at-Risk ($\AVAR$) \cite{rockafellar2002conditional, sarykalin2008value} for the evaluation of the risk of cascading failures.}

{ Our research focuses on examining a platoon of autonomous vehicles that rely on a time-invariant network for communication \cite{ali2015enhanced, tan1998demonstration, verginis2015decentralized}. However, the transmission and processing of information in the vehicles' sensors and actuators experience delays. The platoon is further influenced by the input noise that affects each vehicle independently. This noise represents external perturbations and transforms the system into a stochastic dynamical network \cite{Somarakis2020b}. Our specific focus is on investigating collision events between pairs of consecutive vehicles, particularly when other pairs within the platoon have already experienced systemic failures. Instead of using the Markov Chain \cite{7438924} or flow redistribution model \cite{zhang2018cascading,zhang2019robustness} to model the propagation of the cascading failures in the network, our approach investigate the cascading effect by analyzing the statistical behavior of the steady-state observables, which provides additional insights from a communication graph design perspective. This paper builds upon our previous works on risk of single collision and its boundedness in a vehicle platoon \cite{somarakis2018risk, ghaedsharaf2018performance}, risk of cascading collisions using $\VAR$ \cite{liu2022emergence, liu2021risk}, and presents a significant difference by employing $\AVAR$ as the risk measure. Furthermore, this work considers the {\it cascading} effect from a single existing failure as well as multiple existing failures. We also explore the risk formulation in special graph typologies with certain symmetry and the time-delay induced fundamental limitations on the best achievable risks.}

{ {\it Our Contributions: } 
To the best of our knowledge, this paper introduces the first formal framework for quantifying the risk of cascading failures in large-scale networked control systems with inherent deficiencies. Our theoretical contributions provide a systematic approach for evaluating the resilience of such networks to cascading failures and offer an efficient methodology for validating network design feasibility by assessing the best achievable risk, without the need to exhaustively explore all possible graph topologies. The analysis of the paper are structured as follows: In Section \ref{sec:risk}, we examine the $\AVAR$ of cascading failure by analyzing the steady-state distribution of the vehicle platoon. In Section \ref{sec:specialgraph}, we investigate the contribution and interaction between pairs of vehicles within a specific graph by analyzing the marginal variance and correlation. Section \ref{sec:limits} explores the fundamental limits imposed by time-delays on the steady-state variance of the covariance, enabling us to explore the best achievable risk of cascading collisions under different conditions. Finally, in Section \ref{sec:case} we analyze how the communication graph topology, scale, and distribution of malfunctioning vehicles impact the risk of cascading collisions via extensive simulations.}

We would like to highlight that while this work shares some similarities with the conference papers \cite{liu2021risk,liu2022emergence}, it presents a more thorough and nuanced examination of the effects of inter-vehicle collisions in platoons. This research incorporates novel findings and sections that were not included in the earlier conference versions. Unlike the earlier studies that used the value-at-risk measure, this study adopts the average value-at-risk measure, enhancing our understanding of the risks associated with time-delay in platoon systems. Entirely new sections, such as \ref{subsec:single-event}, \ref{sec:specialgraph}, and \ref{sec:limits}, along with a substantial part of section \ref{sec:case}, are introduced. The theorems and lemmas are articulated using the average value-at-risk concept, significantly reducing their overlap with the content of the previous conference versions.

\section{Preliminaries} 
The $n-$dimensional Euclidean space is denoted by $\R^n$ and $\mathbb{R}^n_{+}$ represents the non-negative orthant of $\R^n$. We denote the vector of all ones by $\bm{1}_n = [1, \dots, 1]^T$. The set of standard Euclidean basis for $\mathbb{R}^{n}$ is represented by $\{\bm{e}_1, \dots, \bm{e}_n\}$ and $\tilde{\bm{e}}_i := \bm{e}_{i+1} - \bm{e}_{i}$ for all $i = 1, \dots, n-1$. 

We consider a simple, connected, undirected, and weighted graph by $\mathcal{G}$, and its corresponding Laplacian matrix of $\mathcal{G}$ is a $n \times n$ matrix $L = [l_{ij}]$ with elements
\[
    l_{ij}:=\begin{cases}
        \; -k_{ij}  &\text{if } \; i \neq j \\
        \; k_{i1} + \ldots + k_{in}  &\text{if } \; i = j 
    \end{cases},
\]
where $k_{ij} \geq 0$ denotes the weight on the link $(i,j)$. The Laplacian matrix of a graph is symmetric and positive semi-definite \cite{van2010graph}, and the smallest Laplacian eigenvalue is zero with algebraic multiplicity one since the graph is connected. The spectrum of $L$ can be ordered as 
$0 = \lambda_1 < \lambda_2 \leq \dots \leq \lambda_n.$ The eigenvector that corresponds to $\lambda_k$ is denoted by $\bm{q}_{k}$. By letting $Q = [\bm{q}_{1} | \dots | \bm{q}_{n}]$, it follows that $L = Q \Lambda Q^T$ with $\Lambda = \text{diag}[0, \lambda_2, \dots, \lambda_n]$. We normalize the Laplacian eigenvectors such that $Q$ becomes an orthogonal matrix, i.e., $Q^T Q = Q Q^T = I_{n}$ with $\bm{q}_1 = \frac{1}{\sqrt{n}} \bm{1}_n$. We refer the detailed introduction of the Algebraic graph theory to the Appendix.

Let $\mathcal{L}^{2}(\mathbb{R}^{q})$ be the set of all $\R^q-$valued random vectors $\bm{z} = [z^{(1)}, \dots ,z^{(q)}]^T$ of a probability space $(\Omega, \mathcal{F}, \mathbb{P})$ with finite second moments. A normal random variable $\bm{y} \in \mathbb{R}^{q}$ with mean $\bm{\mu} \in \mathbb{R}^{q}$ and covariance matrix $\Sigma \in \R^{q \times q}$ is represented by $\bm{y} \sim \mathcal{N}(\bm{\mu}, \Sigma)$. The error function $\text{erf}:\R \rightarrow (-1,1)$ is
$
\text{erf} (x) = \frac{2}{\sqrt{\pi}} \int_{0}^{x} e^{-t^2} \text{d} t,
$
which is invertible on its range as { $\text{erf}^{-1} (x)$  \cite{strecok1968calculation}}. We employ standard notation $\text{d} \bm{\xi}_t$ for the formulation of stochastic differential equations.

\begin{figure}[t]
    \centering
    \includegraphics[width=0.75\linewidth]{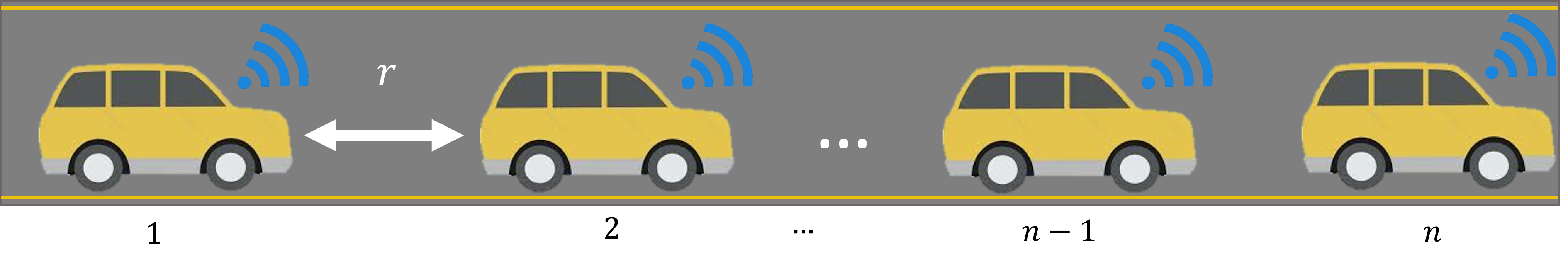}
    \caption{Schematics of platoon ensemble of autonomous vehicles. Speed and distance control is adjusted automatically with feedback laws using information communicated over a virtual network. }
    \label{fig:platoon}
\end{figure}

\section{Problem Statement}

Let us consider a platoon of $n$ vehicles along the horizontal axis as in Fig. \ref{fig:platoon}, which are labeled in a descending order such that the $n$'th vehicle is considered as the leader of the platoon. The state of the $i$'th vehicle is denoted by $[x^{(i)}_{t}, v^{(i)}_{t}]^T$, in which $x^{(i)}_{t}, v^{(i)}_{t} \in \mathbb{R}$ represent the position and velocity of the $i$'th vehicle at time $t$, respectively. The dynamics of the $i$'th vehicle is governed by following stochastic differential equations 
\begin{equation} \label{eq:dyn_vehicle}
    \begin{aligned}
        \textrm{d} x^{(i)}_{t} &= v^{(i)}_{t} \textrm{d} t\\
        \textrm{d} v^{(i)}_{t} &= u^{(i)}_{t} \textrm{d} t + { g_i}\, \textrm{d} \xi^{(i)}_{t}
    \end{aligned}
\end{equation}

\vspace{-2mm}
\noindent where $u^{(i)}_{t} \in \R$ is the control input. We model the uncertainty in the dynamics of each vehicle with ${ g_i}\, \text{d} \xi^{(i)}_{t}$, which represent the white noise generators\footnote{The stochastic process $\xi^{(i)}_{t} \in \mathcal{L}^2(\R)$ denotes the real-valued Brownian motion.}. It is assumed that the white noise is additive and it is independent of other vehicles' noises. The magnitude of the noise is measured by a diffusion parameter ${ g_i} > 0$, and it { might be different} among all vehicles. The objectives of the control of the platoon are to guarantee the following two global behaviors: (i) the { expected} pair-wise differences between position variables $x^{(i)}_{t}$ of every two consecutive vehicles converge to { a target platoon distance $r>0$}; (ii) all vehicles attain the same { expected} velocity in the steady-state. Assuming that for any information to be communicated and processed, there is a non-negligible time-delay\footnote{This assumption has been widely used by other researchers as it allows analytical derivations of formulas. For vehicle platooning, it is a common practice to use identical communication modules for all vehicles, which results in a uniform communication time-delay.}, modeled as a lumped parameter $\tau > 0$. It is known from \cite{yu2010some} that the following feedback control law can achieve objectives (i) and (ii):
\begin{equation*} 
    \begin{aligned}
        u^{(i)}_{t} = \sum_{j=1}^{n} k_{ij} \left( v^{(j)}_{t - \tau} - v^{(i)}_{t - \tau}\right) + \beta \sum_{j=1}^{n} k_{ij} \left( x^{(j)}_{t - \tau} - x^{(i)}_{t - \tau} - (j-i) r \right).
    \end{aligned}
\end{equation*}
The parameter $\beta > 0$ balances the effect of the relative positions and the velocities in the control input and $r>0$ denotes the target platoon distance between two consecutive vehicles. We define the vector of positions, velocities, and noise inputs as $\bm{x}_t = [x_{t}^{(1)}, \dots, x_{t}^{(n)}]^T$, $\bm{v}_t = [v_{t}^{(1)}, \dots, v_{t}^{(n)}]^T$, $\bm{\xi}_t = [\xi_{t}^{(1)}, \dots, \xi_{t}^{(n)}]^T$, and the target distance vector as $\bm{r} = [r,2r,\dots,nr]^T$. The feedback gain $k_{ij} \geq 0$ are designed so that the resulting communication graph stays connected. Then, the closed-loop dynamics can be cast as the following initial value problem
\begin{equation}    \label{eq:dyn}
    \begin{aligned}
        \text{d}\bm{x}_t &= \bm{v}_t \, \text{d} t,\\
        \text{d}\bm{v}_t &= -L \bm{v}_{t-\tau} \,\text{d} t - \beta L (\bm{x}_{t-\tau} - \bm{r}) \,\text{d} t+ { G} \,\text{d} \bm{\xi}_t,
    \end{aligned}
\end{equation}
for all $t \geq 0$ and given deterministic values of $\bm{x}_t$ and $\bm{v}_t$ for $t \in [-\tau, 0]$. { The matrix $G$ is given by $G = \text{diag} \left\{ g_{1}, ..., g_{n}\right\}$.} This poses as stochastic functional differential equations \cite{Evans2013, mohammed1984stochastic}, the standard result guarantees that $\{(\bm{x}_t, \bm{v}_t)\}_{t\geq -\tau}$ is a well-posed stochastic process.

The central {\it problem} studied in this work revolves around evaluating the risk of cascading failures within a platoon of vehicles, particularly in the presence of communication time-delay and exogenous noise. Our primary objective is to quantify an Average Value-at-Risk ($\AVAR$) as a function of the communication graph Laplacian, time-delay, and noise statistics by considering scenarios where a chain of failures (of the same type) has already occurred across different locations within the system. Furthermore, we will explore the explicit formulas of the risk of cascading collisions on specific communication graphs as well as potential inherent limitations on the best achievable levels of risk.

\section{Preliminary Results}   \label{sec:prelims}

{ In this section, we present an overview of several preliminary studies aimed at assessing the stability conditions of the unperturbed system, analyzing the steady-state statistics of key observables, and defining risk measures, which lay the groundwork for the main result.}

\subsection{Internal Stability { of the Unperturbed System}}

{ We investigate stability of the unperturbed closed-loop network, i.e., when $G = 0$ in \eqref{eq:dyn}. Then, we analyze the statistical properties of the solution of \eqref{eq:dyn} when $G \neq 0$.} We denote the convergence of the platoon as the velocities of all vehicles obtain the same value and the distances among all consecutive vehicles are $r > 0$, which is equivalent to 
\begin{equation*}
    \lim_{t \rightarrow \infty} |v_{t}^{(i)} - v_{t}^{(j)}| = 0 ~\text{and}~ \lim_{t \rightarrow \infty} |x_{t}^{(i)} - x_{t}^{(j)} - (i-j) r| = 0
\end{equation*}
for all $i,j = 1,...,n$. Recent works \cite{Somarakis2020b,yu2010some} have shown that the deterministic platoon will converge if and only if 
\begin{align}   \label{eq:stable}
    \Big(\lambda_i \, \tau, \beta \, \tau \Big) \in S,
\end{align}
for all $i=2,...,n$, and we consider the set
\begin{align*}
    S = \bigg\{(s_1,s_2) \in \R^2 \bigg| s_1 \in \bigg(0, \frac{\pi}{2} \bigg), s_2 \in \left(0, \frac{a}{\tan(a)} \right), \text{ for } a \in \left(0,\frac{\pi}{2} \right) \text{ the solutions of } a\sin(a) = s_1 \bigg\}.
\end{align*} 
In this work, we only consider the cases when the deterministic platoon dynamic is convergent, i.e., $(\lambda_i \tau, \beta \tau ) \in S$, where $\lambda_i$ denotes the eigenvalues of the graph Laplacian matrix $L$.

\subsection{Steady-State Statistics of Inter-vehicle Distances}
Using the decomposition of $L = Q \Lambda Q^T$,  the dynamics of the platoon \eqref{eq:dyn} can be mapped into another coordinate by the state transformation
$$
    \bm{z}_t = Q^T (\bm{x}_t -\bm{r}) \text{, and } \bm{\textup{v}}_t = Q^T \bm{v}_t.
$$
 The platoon system in the new coordinate is given by 
\begin{equation}\label{eq:dyn_z}
    \begin{aligned}
        \text{d}\bm{z}_t &= \bm{\textup{v}}_t \, \text{d}t,\\
        \text{d}\bm{\textup{v}}_t &= -\Lambda \bm{\textup{v}}_{t-\tau} \, \text{d}t \; - \; \beta \Lambda \bm{z}_{t-\tau} \, \text{d}t \;+\;   Q^T \, { G} \, \text{d} \bm{\xi}_t.
    \end{aligned}
\end{equation}
Once the stability condition of \eqref{eq:dyn_z} is satisfied, the solution of decoupled subsystems can be represented as
\begin{equation}
    \begin{aligned}
        \begin{bmatrix}
        z_t^{(i)} \\
        \textup{v}_t^{(i)} 
        \end{bmatrix} = 
       \Gamma(z_{[-\tau,0]};\textup{v}_{[-\tau,0]}) + \int^{t}_{0} \Phi_i(t-s) B_i \, { G} \, \text{d} \bm{\xi}_s,
    \end{aligned}
\end{equation}
for $\Gamma(\cdot \,; \cdot )$ a generalized function describing transient dynamics that vanish exponentially fast \cite{Somarakis2020b}, { $\Phi_i(t)$ the principal solution of the unperturbed system,} and $B_i = [\bm{0}_{1 \times n}, \bm{q}_i^T]^T$. It is known from \cite{Somarakis2020b} that the statistics of the steady-state $\bar{\bm{z}} \in \R^{n}$ follows a multi-variate normal distribution with mean $\bm{0}$ and the covariance matrix is shown by
$
    \Sigma_z = \text{diag} \left\{ \sigma_{z_1}^2, ..., \sigma_{z_n}^2\right\},
$
in which $\sigma_{z_i}^2 = \frac{ { g_i^2} \tau^3}{2\pi}$ $ f(\lambda_i\tau, \beta \tau)$ with 
\begin{equation} \label{eq:f}
    \begin{aligned}
        f(s_1, s_2) = \int_{\R} \frac{\text{d}\,r}{(s_1s_2 - r^2 \cos(r))^2 + r^2 (s_1-r \sin(r))^2}.
    \end{aligned}
\end{equation}

Given that $\bm{x}_{t} = Q \bm{z}_{t} + \bm{y}$, we define the steady-state distance vector of the platoon as $\bar{\bm{d}} \in \mathbb{R}^{n-1}$ such that
\begin{align}   \label{eq:dist}
    \bm{\bar{d}} = D Q \bm{\bar{z}} + r \bm{1}_n,
\end{align}
in which $D = \left[\tilde{\bm{e}}_1^T \big| \dots \big|\tilde{\bm{e}}_{n-1}^T \right]^T$ and $Q = [ \bm{q}_1 | \dots | \bm{q}_n]$. The $i$'th element in $\bm{\bar{d}}$ denotes the distance between the $i$'th and $(i+1)$'th vehicle in the platoon.
 
\begin{lemma}     \label{lem:d_steady}
    Suppose the stability condition \eqref{eq:stable} holds, the steady-state inter-vehicle distance vector $\bar{\bm{d}}$ follows a multi-variate normal distribution in $\R^{n-1}$ such that
    \begin{align*}
        \bm{\bar{d}} \sim \mathcal{N} \Big(r \bm{1}_{n-1}, \, \Sigma \Big),
    \end{align*}
    with the elements of $\Sigma = [\sigma_{ij}]$ is shown by
    \begin{align} \label{eq:sigma_d}
        \sigma_{ij} = 
        \frac{\tau^3}{2\pi} \sum_{k=2}^{n}  \big(\tilde{\bm{e}}_{i}^T \bm{q}_k \big) \big(\tilde{\bm{e}}_{j}^T \bm{q}_k \big) { \, g_k^2 \,} f(\lambda_k \tau, \beta \tau),
    \end{align}
    for all $i,j = 1,\dots, n-1$, and $f(\cdot, \cdot)$ as in \eqref{eq:f}. We also denote the diagonal elements $\sigma_{ii}$ as $\sigi^2$. 
\end{lemma}

\begin{remark}
    In the case when time-delay $\tau = 0$, the expression of \eqref{eq:sigma_d} boils down to 
    \begin{equation*}
        \sigma_{ij} = 
        \frac{1}{2\pi} \sum_{k=2}^{n} \big(\tilde{\bm{e}}_{i}^T \bm{q}_k \big) \big(\tilde{\bm{e}}_{j}^T \bm{q}_k \big) { g_k^2} \int_{\R} \frac{\textup{d} \, r}{(\lambda_k \beta - r^2)^2+r^2\lambda_k^2}.
    \end{equation*}
\end{remark}

Upon observing \eqref{eq:sigma_d}, { when the external noise magnitude is identical among all vehicles, i.e.,  $g_1=...=g_n=g$,} it becomes evident that the cross-correlation $\rho_{ij} = \sigma_{ij}/\sigi\sigj$ remains unaffected by the magnitude of external noise, which is denoted as $g$. Consequently, it can be inferred that the covariance between the relative distances of two pairs of vehicles depend on the time-delay and characteristic of the underlying communication graph.

\subsection{Risk Measures}      \label{sec:risk_measure}
To assess the level of uncertainty inherent in the relative distances between vehicles, we utilize the concept of Value-at-Risk ($\VAR$), and Average Value-at-Risk ($\AVAR$) \cite{Follmer2016,rockafellar2002conditional,sarykalin2008value}. Both risk measures quantify the likelihood of a random variable falling within an undesirable range of values, such as a near-collision scenario. The set of undesirable values is known as an unsafe set and is denoted as $C \subset \R$. In the probability space $(\Omega, \mathcal{F}, \mathbb{P})$, the unsafe events of the random variable $y: \Omega \rightarrow \R$ are defined as $\{ \omega \in \Omega ~|~y(\omega) \in C\}$. 

In practice, one can design and tailor level sets to cover a suitable neighborhood of $C$ to characterize alarm zones as a random variable approaches $C$. We establish a collection of level sets $\{C_{\delta}~|~\delta \in [0,\infty]\}$ of $C$ that satisfy the following conditions for any sequence $\{\delta_n\}^{\infty}_{n=1}$ with property $\lim_{n \rightarrow \infty} \delta_n = \infty$:

\begin{equation}    \label{eq:level_set}
    \begin{aligned}
        \hspace{-4cm} 1) \hspace{2mm} &C_{\delta_1} \subset C_{\delta_2} \text{ when } \delta_1 > \delta_2\\
        \hspace{-4cm} 2) \hspace{2mm} &\lim_{n \rightarrow \infty} C_{\delta_n} = \bigcap_{n=1}^{\infty} C_{\delta_n} = C.
    \end{aligned}
\end{equation}

\vspace{-2mm}
The Value-At-Risk $\VAR$ of a continuous random variable $y$ is defined as
\footnote{Unlike the conventional definition of $\mathfrak{R}_\varepsilon := \inf \left\{ z \, | \, \mathbb{P} \left\{ y > z  \right\} < \varepsilon \right\}$, we consider the other tail of the distribution in the view of collision events.}
\begin{equation}
    \mathfrak{R}_\varepsilon := \inf \left\{ z \, \big| \, \mathbb{P} \left\{ y < z  \right\} > \varepsilon \right\},
\end{equation}
and the corresponding $\AVAR$ \cite{rockafellar2002conditional} is given by 
\begin{equation}
    \mathfrak{A}_\varepsilon := \mathbb{E} \left[ y \, \big| \, y < \mathfrak{R}_\varepsilon \right],
\end{equation}
where $\varepsilon \in (0,1)$ denotes the confidence level. The value of $\sup \{ \delta \geq 0 \,\big|\, \mathfrak{A}_\varepsilon \in C_{\delta}\}$ indicates how dangerously $y$ can get to $C$ as well as the expected severity when such event occurs. Following this notation, the representation of $\AVAR$ in terms of the parameter $\delta$ of level sets $C_\delta$ is defined as follows:
\begin{equation}
    \mathcal{A}_\varepsilon := \sup \left\{ \delta \geq 0 \, \Big| \, \mathfrak{A}_{\varepsilon} \in C_\delta  \right\},
\end{equation}
where a smaller $\varepsilon$ indicates a higher level of confidence on random variable $y$ to stay away from $C$. The case $\mathcal{A}_{\varepsilon}=0$ signifies that $\mathbb{E} \left[ y \, | \, y < \mathfrak{R}_\varepsilon \right]$ lands on the boundary or outside of $C_0$. We have $\mathcal{A}_{\varepsilon} > 0$ if and only if $\mathbb{E} \left[ y \, | \, y < \mathfrak{R}_\varepsilon \right] \in C_{\delta}$ for some $\delta >0$. The extreme case $\mathcal{A}_{\varepsilon}=\infty$ indicates that $\mathbb{E} \left[ y \, | \, y < \mathfrak{R}_\varepsilon \right]$ is to be found inside $C$. Unlike the conventional definition of $\AVAR$ which is a coherent risk measure \cite{sarykalin2008value}, the term $\mathcal{A}_\varepsilon$ only preserves monotonicity and sub-additivity.

\section{Risk of Cascading Collisions}      \label{sec:risk}

Let us consider that the platoon ensemble travels in a straight line, as illustrated in Fig. \ref{fig:platoon}. Naturally, there is a sequential enumeration ${1,\dots,n-1}$ for pairs of consecutive vehicles. Following the notation of Lemma \ref{lem:bivariate}, let $i$ and $j$ be two pairs of vehicles in the ensemble, where $i$'th pair means vehicles $i$ and $i+1$ and so forth. 

\begin{figure}[t]
    \centering
    \includegraphics[width=0.75\linewidth]{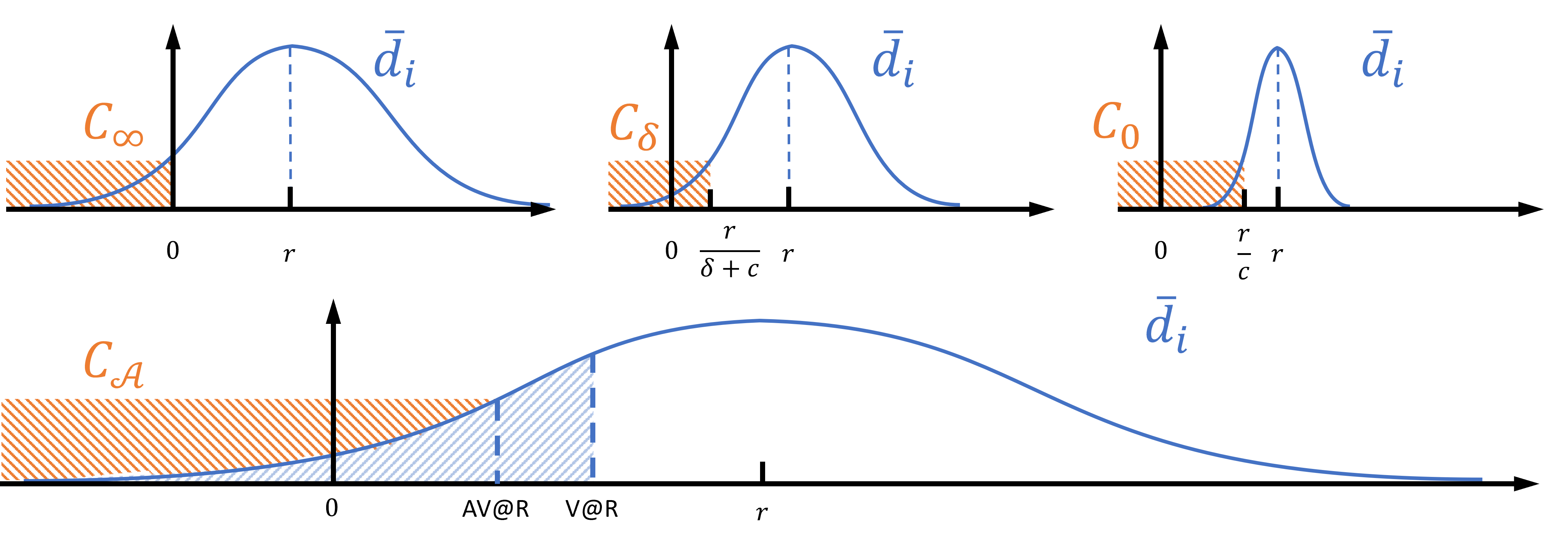}
    \caption{Change of the set $C_{\delta}$ with different distributions of the inter-vehicle distance. }
    \label{fig:unsafe_set}
\end{figure}

The occurrence of an inter-vehicle collision in the $i$'th pair is defined by the event ${\bar{d}_i \in (-\infty,0) = C}$, and its corresponding level set can be designed as follows:
\begin{align}
    C_{\delta}=\Big( -\infty, h(\delta) \Big),
\end{align}
where $\delta \in [0,\infty]$. The selection of function $h: [0,\infty] \rightarrow \R_+$ should ensure that $C_\delta$ satisfies the properties outlined in \eqref{eq:level_set}. In this paper, we adopt the function $h(\delta) = \frac{r}{\delta + c}$, resulting in the following expression for the level set:
\begin{align}
    C_{\delta}=\left( -\infty, \frac{r}{\delta+c} \right),
\end{align}
where $c \geq 1$, and the visual illustration of $C_{\delta}$ is shown in Fig. \ref{fig:unsafe_set}. Let us discuss scenarios that exemplify a few special cases of cascading collisions based on the given information on the occurred collisions.

\subsection{Range Information of the Single Vehicle Pair}\label{subsec:single-event}
Let us consider the situation when one has obtained the range information about the current distance of the $i$'th pair, represented by $\bar{d}_i \in C_{\delta^*}$, where $C_{\delta^*} = (-\infty, \frac{r}{\delta^*+c})$. An example is when a pair of vehicles is known to be dangerously close to collision, but their exact inter-vehicle distance is unknown due to the limited measurement capability, denoted as $\bar{d}_i \in C_{\delta^*}$ with a sufficiently large $\delta^*$. 
In such scenarios, the risk of cascading collisions at the $j$'th pair is defined as:
\vspace{-2mm}
\begin{equation}
    \hspace{-2mm}\mathcal{A}^{i,j}_{\varepsilon} = \sup \left\{ \delta \geq 0 \,\Big|\, \mathbb{E} \left[ \bar{d}_j \, \big| \, \bar{d}_j  < \mathfrak{R}_\varepsilon \wedge \bar{d}_i \in C_{\delta^*} \right] \in C_\delta \right\}, \hspace{-2mm}
\end{equation}
\vspace{-2mm}
with
\vspace{-2mm}
\begin{equation}
    \mathfrak{R}_\varepsilon := \inf \left\{ z \, \Big| \, \mathbb{P} \left\{\bar{d}_j < z \, \big | \, \bar{d}_i \in C_{\delta^*} \right\} > \varepsilon \right\}.
\end{equation}

\begin{theorem}   \label{thm:risk_levelset}
    Suppose that the platoon \eqref{eq:dyn} reaches the steady-state and the distance of the $i$'th pair is observed to be $\bar{d}_i \in C_{\delta^*}$. The risk of cascading collision $\cass$ at the $j$'th pair is 
    \begin{equation}
    \begin{aligned}
        \cass:=\begin{cases}
            0, &\text{if} ~ \mathfrak{A}^{i,j}_\varepsilon \geq \frac{d}{c}\\[5pt]
            \dfrac{r}{\mathfrak{A}^{i,j}_\varepsilon} - c, &\text{if} ~ 0 < \mathfrak{A}^{i,j}_\varepsilon < \frac{d}{c} \\[5pt]
            \infty, &\text{if} ~ \mathfrak{A}^{i,j}_\varepsilon \leq 0
            \end{cases},
    \end{aligned}
    \end{equation}
    where the closed-form representation of $\mathfrak{A}^{i,j}_\varepsilon$ is shown in the Appendix.
\end{theorem}

The above result presents the risk of cascading collision when a pair of vehicles is identified as being dangerously close to inter-vehicle collision. It is important to note that, apart from collisions, this result can be utilized to quantify the risk of cascading failures in various scenarios, e.g., vehicles detachment \cite{Somarakis2020b}, provided that the level set $C_\delta$ is appropriately defined. Due to the lack of \textit{exact information}, the current analysis utilizes the \textit{best available information} and it is constrained by the implicit structure of the expression. One can deploy state estimation techniques, e.g., Kalman filter, to get a better idea of $\bar{d}_i$, but it is beyond the scopes of this work. Therefore, to facilitate further analysis, we will focus on a more practical and specific case in the remaining sections of this paper, which allows for a more comprehensive examination of cascading behavior.

\subsection{Snapshot of a Single Vehicle Pair}
We now consider a special case where a single, isolated collision affects a pair of vehicles, e.g., the $i$'th pair, and we have access to the measurement of its inter-vehicle distance $\bar{d}_i$ with full precision. How does such an event affect the rest of the network? The answer follows from the closed-form expression for the statistical analysis of the steady-state distance, denoted as $\bar{d}_j$, when the event has occurred at the $i$'th pair.

Let us assume the distance of the $i$'th pair is measured as $\bar{d}_i = d^*$, and we study the risk of cascading collision at the $j$'th pair in the view of the previous observation. The event of the cascading collision is defined as 
\begin{equation}
    \left\{\bar{d}_{j} \in C \, \big| \, \bar{d}_i = d^* \right\},
\end{equation}
where $d^* \geq 0$. When $d^* = 0$, it signifies an ongoing collision at the $i$'th pair. The associated risk of the cascading collision is then defined as follows:
\begin{align}       \label{eq:def_avar_single}
    \hspace{-2mm} \mathcal{A}^{i,j}_{\varepsilon} = \sup \left\{ \delta \geq 0 \,\Big|\, \mathbb{E} \left[ \bar{d}_j \, \big| \, \bar{d}_j < \mathfrak{R}_\varepsilon^{i,j} \wedge \bar{d}_i = d^* \right] \in C_\delta \right\}, \hspace{-2mm}
\end{align}
where
{ \begin{align}   
    \mathfrak{R}^{i,j}_{\varepsilon} = \inf \left\{ z \,\Big|\, \mathbb{P} \left\{ \bar{d}_{j} < z \,\big|\, \bar{d}_i = d^* \right\} > \varepsilon \right\},
\end{align}}
with the confidence level $\varepsilon \in (0,1)$. The larger value of $\mathcal{A}^{i,j}_{\varepsilon}$ indicates a higher probability that the cascading collision is going to occur to the system and the collision is likely to become more severe. Given the knowledge of the existing event at $i$, the conditional statistics at the $j$'th pair is given as follows.

\begin{lemma}   \label{lem:bivariate}
    For any pair of two steady-state distances $\bar{d}_i$ and $\bar{d}_{j}$, the conditional distribution of $\bar{d}_{j}\,|\,\bar{d}_{i} = d^*$ follows a normal distribution $\mathcal{N}(\tilde{\mu},\tilde{\sigma}^2)$, where
    \begin{equation}    \label{eq:single_cond_stat}
        \tilde{\mu} = r + \rho_{ij} \frac{\sigj}{\sigi}(d^* - r),\text{ and }~\tilde{\sigma}^2 = \sigj^2 (1 - \rho_{ij}^2),
    \end{equation}
    in which $\rho_{ij} = \sigma_{ij} / \sigma_{i} \sigma_{j}$ and $|\rho_{ij}| < 1$. The values of $\sigma_{ij}, \sigma_{i}, \text{ and }\sigma_{j}$ are computed using \eqref{eq:sigma_d}.
\end{lemma}

Then, the risk of cascading collision $\cass$ is presented in the following result.

\begin{theorem}     \label{thm:avar_single_cas}
     Suppose that the platoon \eqref{eq:dyn} reaches the steady-state and the $i$'th pair is observed with measurement $\bar{d}_i = d^*$. The risk of cascading collision at the $j$'th pair is 
     \begin{equation}
         \begin{aligned}
             \mathcal{A}_{\varepsilon}^{i,j} :=\begin{cases}
            0, &\text{if} ~ \frac{c\tilde{\mu} - r}{c \tilde{\sigma}} \geq \kappa_{\varepsilon}\\
            \dfrac{r}{\tilde{\mu}-\kappa_\varepsilon \tilde{\sigma}} - c, &\text{if} ~ \kappa_{\varepsilon} \in \big(\frac{c\tilde{\mu}-r}{c \tilde{\sigma}}, \frac{\tilde{\mu}}{\tilde{\sigma}}  \big) \\
            \infty, &\text{if} ~ \frac{\tilde{\mu}}{\tilde{\sigma}} \leq  \kappa_{\varepsilon}
            \end{cases},
         \end{aligned}
     \end{equation}
    where $\kappa_{\varepsilon} = \left(\sqrt{2\pi} \varepsilon \exp(\io^2) \right)^{-1}$, $\io = \textup{erf}^{-1} (2\varepsilon-1)$, $\tilde{\mu}$ and $\tilde{\sigma}$ are the conditional statistics computed using \eqref{eq:single_cond_stat}.
\end{theorem}

\begin{figure}[t]
    \centering
    \includegraphics[width=0.75\linewidth]{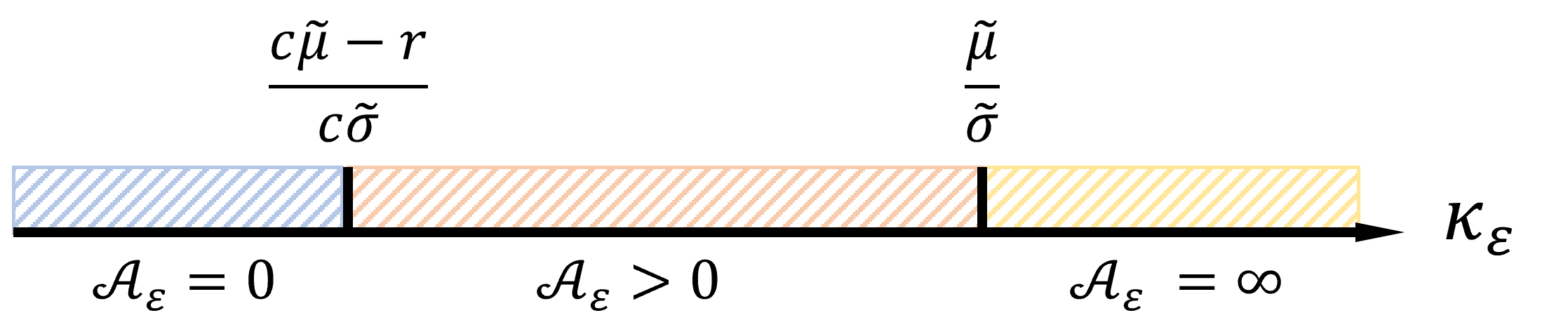}
    \caption{The risk partition on the axis of $\kappa_\varepsilon = (\sqrt{2\pi} \varepsilon\exp(\io^2))^{-1}$.}
    \label{fig:kappa}
\end{figure}

The above theorem presents the risk of cascading collision with branches determined by the conditional statistics of $\bar{d}_i$ and the confidence level embedded in $\kappa_{\varepsilon}$. As shown in Fig. \ref{fig:kappa}, there are two extreme scenarios to consider. Firstly, if $\frac{c\tilde{\mu}-r}{c \tilde{\sigma}} \geq \kappa_{\varepsilon}$, there does not exists a $\delta \geq 0$ such that \eqref{eq:def_avar_single} will be satisfied. Conversely, if $\frac{\tilde{\mu}}{\tilde{\sigma}} \leq \kappa_{\varepsilon}$, the conditional expectation inside \eqref{eq:def_avar_single} will land outside the level set $C_\delta$ for any given $\delta \geq 0$. In any other case, the risk of cascading collision will assume a positive finite value, indicating how dangerously the $\AVAR$ of pair $j$ is close to the collision.

We form the risk profile vector $\bm{\mathcal{A}}^{i}_{\varepsilon} \in \R^{n-2}$ of the platoon by stacking up all risks of cascading collision among remaining pairs, such that
\begin{equation}
    \bm{\mathcal{A}}^{i}_{\varepsilon} = \big[\mathcal{A}^{i,1}_{\varepsilon}, \dots, \mathcal{A}^{i,i-1}_{\varepsilon}, \mathcal{A}^{i,i+1}_{\varepsilon}, \dots , \mathcal{A}^{i,n-1}_{\varepsilon} \big]^T.
\end{equation}

\vspace{2mm}

\begin{figure}[t]
    \centering
    \includegraphics[width=0.75\linewidth]{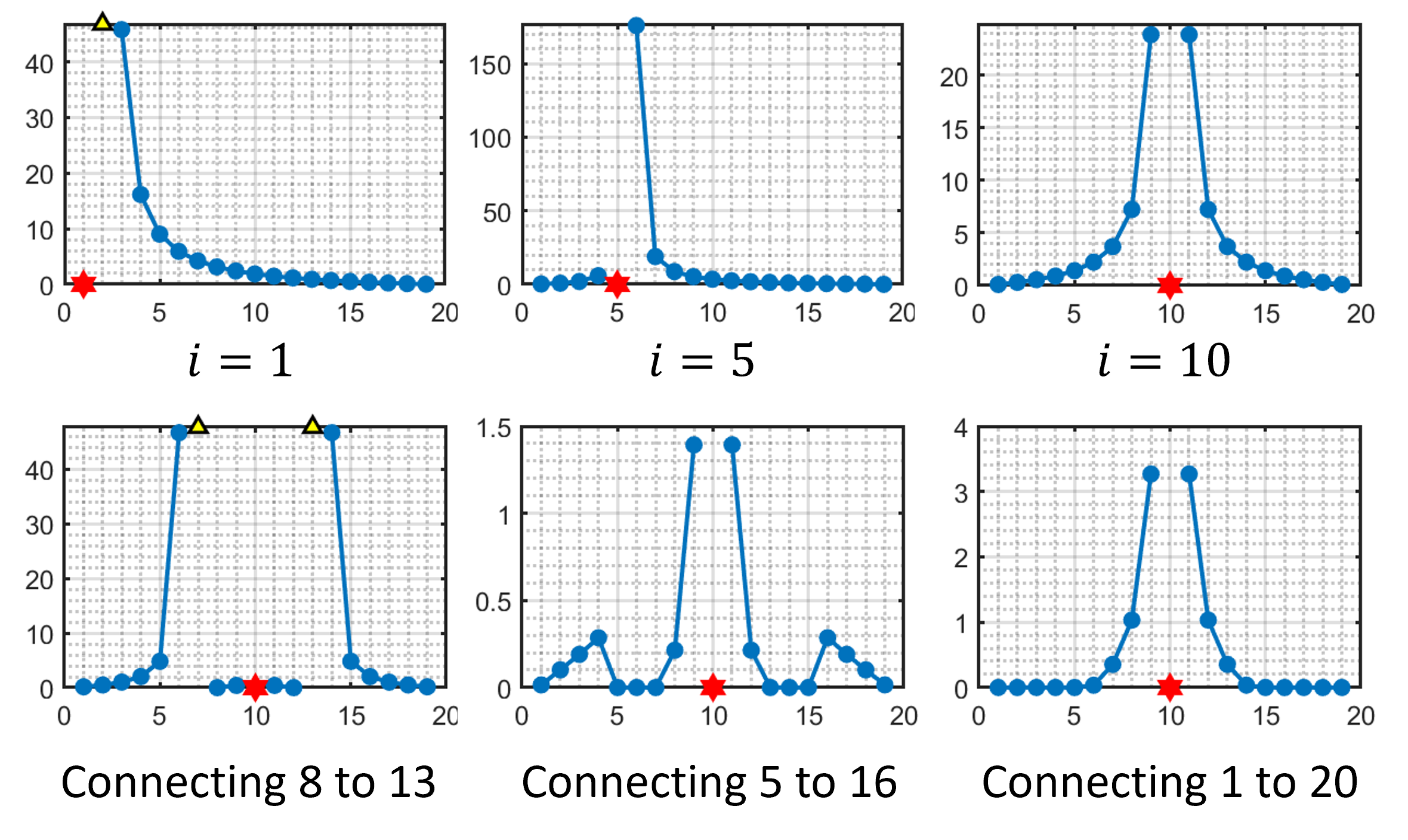}
    \caption{{The risk of cascading collision $\cass$ in a path graph. The x-axis denotes the vehicle number and the y-axis represents the risk value $\cass$. The second row shows the risk profile after adding an edge to the path graph. The existing collision is shown by the red hexagram and the case $\cass = \infty$ is shown by the yellow triangle.}}
    \label{fig:exmaple_1}
\end{figure}

\begin{figure}[t]
    \centering
	\includegraphics[width=\linewidth]{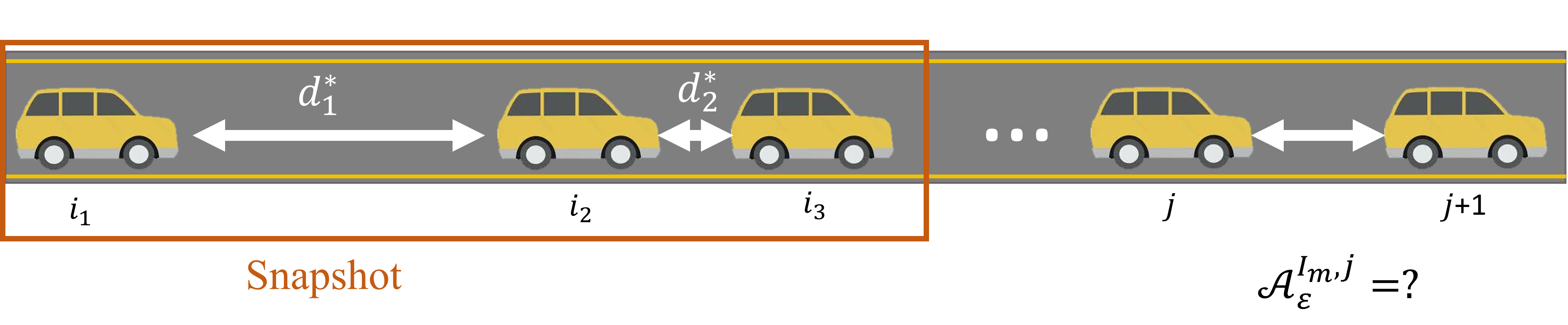}
	\caption{The above figure illustrates the concept of snapshots of multiple vehicle pairs within a platoon.}
	\label{fig:snapshot}
\end{figure}

\noindent {\bf Example 1:}
By using a path communication graph, we consider a vehicle platoon scenario on the highway, where vehicles are capable of communicating solely with their front and rear neighbors. { We set $g_1 =...= g_n = 0.1, \tau = 0.04, \text{ and } \beta = 1$} and assume one pair of vehicles has collided, i.e., $d^* = 0$, with the result shown in Fig. \ref{fig:exmaple_1}. The risk profile reveals that the risk of cascading collisions $\cass$ tends to increase when the vehicles in the network gets closer to the existing collision. The results also demonstrate the risk profile when introducing an additional edge into the existing communication graph. Our findings suggest that by adding a communication link of moderate length, it becomes possible to effectively mitigate the risk of cascading collisions within the platoon.


\subsection{Snapshots of Multiple Vehicle Pairs}  \label{sec:multi_failure}

The following framework enables us to consider and evaluate risk of cascading collisions when more than one single occurred collision is known.
  
Our goal is to quantify the risk of cascading collisions by extending Lemma \ref{lem:bivariate} with utilizing the knowledge of arbitrary number of existing events. We assume that there are $m < n$ observed events in the network, labeled as $\mathcal{I}_m ={ i_1, \cdots, i_m}$. Specifically, we aim to evaluate the risk of cascading collision at the $j$'th pair with $j \notin \mathcal{I}_m$. The distances of $\mathcal{I}_m$ are represented by
\begin{equation}
    \bm{\bar{d}}_{\mathcal{I}_m}=[\bar{d}_{i_1}, ... , \bar{d}_{i_m}]^T  = \bm{d}^*,
\end{equation}
where $\bm{d}^* = [d^*_1, ... , d^*_m]^T$ corresponds to the current measurements, which can be interpreted as taking a partial snapshot of a platoon running on the highway, as depicted in Fig. \ref{fig:snapshot}. For example, $d^*_1 = 0$ characterizes the existence of an inter-vehicle collision happening at the $i_1$'th pair. Then, the risk of cascading collisions at the $j$'th pair is defined as
\begin{equation}
    \begin{aligned}   
        \mathcal{A}^{\mathcal{I}_m,j}_{\varepsilon} = \sup \left\{ \delta \geq 0 \,\Big|\, \mathbb{E} \left[ \bar{d}_j \, \big| \, \bar{d}_j < \mathfrak{R}^{\mathcal{I}_m,j}_{\varepsilon} \wedge \bar{\bm{d}}_{\mathcal{I}_m} = \bm{d}^* \right] \in C_\delta \right\},  
    \end{aligned}
\end{equation}
where
{ \begin{align}   
    \mathfrak{R}^{\mathcal{I}_m,j}_{\varepsilon} = \inf \left\{ z \,\Big|\, \mathbb{P} \left\{ \bar{d}_{j} < z \,\big|\, \bar{\bm{d}}_{\mathcal{I}_m} = \bm{d}^* \right\} > \varepsilon \right\}
\end{align}}
with the confidence level $\varepsilon \in (0,1)$. To evaluate the conditional statistics, we will consider the following block covariance matrix 
\begin{equation}    \label{eq:new-sig}
    \tilde{\Sigma} = \begin{bmatrix}\,
       \tilde{\Sigma}_{11} &\tilde{\Sigma}_{12}\\
       \tilde{\Sigma}_{21} &\tilde{\Sigma}_{22}\,
    \end{bmatrix},
\end{equation}
where
$
\tilde{\Sigma}_{11} = \sigj^2, \, \tilde{\Sigma}_{12} = \tilde{\Sigma}_{21}^T = [\sigma_{j i_1},...,\sigma_{j i_m}], \, \tilde{\Sigma}_{22} = [\sigma_{k_1 k_2}] \in \R^{m \times m}$ for all $k_1, k_2 \in \mathcal{I}_m$ according to \eqref{eq:sigma_d}. Then, the conditional distribution of $\bar{d}_{j} ~\big|~ \bm{\bar{d}}_{\mathcal{I}_m} = \bm{d}^*$ are shown in the following result.

\begin{lemma}   \label{lem:multi_condition}
     Suppose that $\bm{\bar{d}} \sim \mathcal{N}(r \bm{1}_{n-1}, \Sigma)$, the conditional distribution of $\bar{d}_j$ given $\bm{\bar{d}}_{\mathcal{I}_m} = \bm{d}^*$ follows a normal distribution  $\mathcal{N}(\tilde{\mu}, \tilde{\sigma})$, where
    \begin{equation*}
        \tilde{\mu} = r - \tilde{\Sigma}_{12} \, \tilde{\Sigma}_{22}^{-1}(\bm{d}^* -  r\bm{1}_m),
        \text{ and }
        \tilde{\sigma}^2 = \tilde{\Sigma}_{11} - \tilde{\Sigma}_{12} \, \tilde{\Sigma}_{22}^{-1} \, \tilde{\Sigma}_{21}.
    \end{equation*}
\end{lemma}

Then, the risk of cascading collision at the $j$'th pair with measurements of multiple vehicle pairs is presented as follows.

\begin{corollary}   \label{cor:risk_mul_cas}
    Suppose that the platoon \eqref{eq:dyn} reaches the steady-state and pairs with labels $i_1, \cdots, i_m$ are measured as $\bm{\bar{d}}_{\mathcal{I}_m} = \bm{d}^*$. The risk of cascading collision at the $j$'th pair is 
    \begin{equation}
        \begin{aligned}
            \mathcal{A}_{\varepsilon}^{\mathcal{I}_m,j} :=\begin{cases}
            0, &\text{if} ~ \frac{c\tilde{\mu} - r}{c \tilde{\sigma}} \geq \kappa_{\varepsilon}\\
            \dfrac{r}{\gamma(\tilde{\Sigma}, \bm{d}^*, \varepsilon)} - c, &\text{if} ~ \kappa_{\varepsilon} \in \big(\frac{c\tilde{\mu}-r}{c \tilde{\sigma}}, \frac{\tilde{\mu}}{\tilde{\sigma}}  \big) \\
            \infty, &\text{if} ~ \frac{\tilde{\mu}}{\tilde{\sigma}} \leq  \kappa_{\varepsilon}
            \end{cases},
        \end{aligned}
    \end{equation}
    where
    \begin{equation*}
            \gamma(\tilde{\Sigma}, \bm{d}^*, \varepsilon) = r - \tilde{\Sigma}_{12}  \tilde{\Sigma}_{22}^{-1}(\bm{d}^* -  r\bm{1}_m) - \kappa_\varepsilon \sqrt{\tilde{\Sigma}_{11} - \tilde{\Sigma}_{12} \tilde{\Sigma}_{22}^{-1}  \tilde{\Sigma}_{21}},
    \end{equation*}
    $\tilde{\mu}$ and $\tilde{\sigma}$ are calculated as in Lemma \ref{lem:multi_condition}, $\kappa_{\varepsilon} = \left(\sqrt{2\pi} \varepsilon \exp(\io^2) \right)^{-1}$, and $\io = \textup{erf}^{-1} (2\varepsilon-1)$.
\end{corollary}

The risk of a cascading collision with multiple existing events $\casm$ follows a similar structure to $\cass$ despite the calculation of the mean and variance of the conditional distribution differs. It is important to note that the measurement of $\mathcal{I}_m$ can take various forms. For example, in the case of multiple collisions in the platoon, one can assess the risk of cascading collisions by assuming $\bm{\bar{d}}_{\mathcal{I}_m} = \bm{0}$. Alternatively, when given a partial snapshot of the platoon, the risk of collision at certain pairs can be further evaluated with this additional information. In such cases, the measurement of pairs $\mathcal{I}_m$ {\it do not} need to be identical.

Let us also introduce the risk profile vector $\bm{\mathcal{A}}^{\mathcal{I}_m}_{\varepsilon} \in \R^{n-1}$ as follows
\begin{equation}
    \bm{\mathcal{A}}^{\mathcal{I}_m}_{\varepsilon} = \big[\mathcal{A}^{\mathcal{I}_m,1}_{\varepsilon}, \dots, \mathcal{A}^{\mathcal{I}_m,j}_{\varepsilon}, \dots,\mathcal{A}^{\mathcal{I}_m,n-1}_{\varepsilon} \big]^T,
\end{equation}
in which $\mathcal{A}^{\mathcal{I}_m,j}_{\varepsilon} = 0$ if $j \in \mathcal{I}_m$.

\begin{figure}[t]
    \centering
    \includegraphics[width=0.75\linewidth]{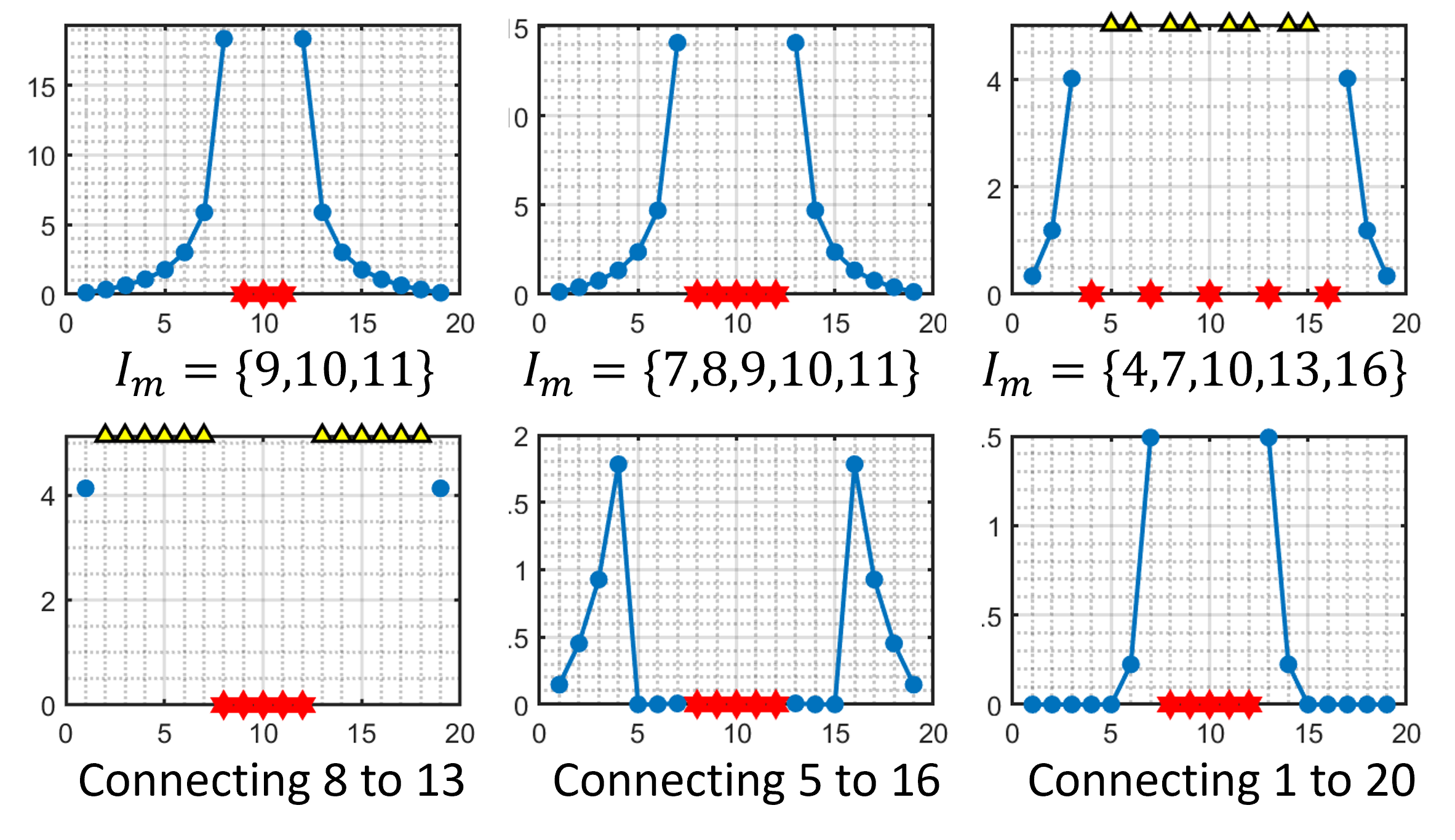}
    \caption{{ The risk of cascading collision $\casm$. The x-axis denotes the vehicle index and the y-axis represents the risk value $\cass$. The second row shows the risk profile after adding an edge to the path graph. The existing collision is shown by the red hexagram and the case $\casm = \infty$ is shown by the yellow triangle}}
    \label{fig:exmaple_2}
\end{figure}

\vspace{2mm}

\noindent {\bf Example 2:} 
We conducted experiments on a path graph with { $g_1 = ... = g_n = 0.1$, $\tau = 0.04$, and $\beta = 1$}, assuming collision occurrences between certain vehicle pairs, denoted by $\bar{\bm{d}} = \bm{0}$. The obtained results are presented in Fig. \ref{fig:exmaple_2}. Our analysis indicates that as the number of existing failures increases, $\casm$ tends to decrease. Additionally, the system becomes more vulnerable when the existing collisions are sparsely distributed in the platoon. Moreover, our investigation into the impact of adding an additional edge to the communication graph reveals that increasing connectivity does not always yield benefits. Instead, introducing a long-distance connection is a more favorable approach for mitigating the risk of cascading collisions.

\section{Risk of Cascading Collisions on Special Graph Topologies}  \label{sec:specialgraph}

The interconnection topology of the underlying communication graph plays a crucial role in determining the extent to which uncertainties propagate in networked systems \cite{siami2016fundamental, Somarakis2020b}. In this section, we explore various graph topologies characterized by specific symmetries { and assuming $g = g_1,...,g_n$ }to gain insights into cascading failures in vehicle platooning.

\subsection{Complete Graph}
For a platoon of vehicles \eqref{eq:dyn} adopting a unweighted complete communication graph, the eigenvalues of the corresponding Laplacian matrix are: $\lambda_1 = 0$ and $\lambda_j = n$ for all $j = 2,\dots,n$. One can find the closed-form statistics of the system's observables as the following.

\begin{lemma}   \label{lem:sig_complete}
    For a platoon \eqref{eq:dyn} with a complete communication graph, the steady-state distance follows $\bm{\bar{d}} \sim \mathcal{N}(r \bm{1}_{n-1}, \Sigma)$, and the covariance matrix is given element-wise by
    \begin{equation}
        \begin{aligned}
            \sigma_{ij}: = 
            \begin{cases}
                \sigma_c,               &\text{ if} ~ i = j\\[2pt]
                -\dfrac{\sigma_c}{2},   &\text{ if} ~ |i-j| = 1\\[2pt]
                0,                      &\text{ if} ~ |i-j| > 1
            \end{cases},
        \end{aligned}
    \end{equation}
     where $\sigma_c = \dfrac{g^2 \tau^3 f(n \tau, \beta \tau)}{\pi}$ for all $i, j = 1,\dots,n-1$.
\end{lemma}

\begin{figure}[t]
    \centering
	\includegraphics[width=\linewidth]{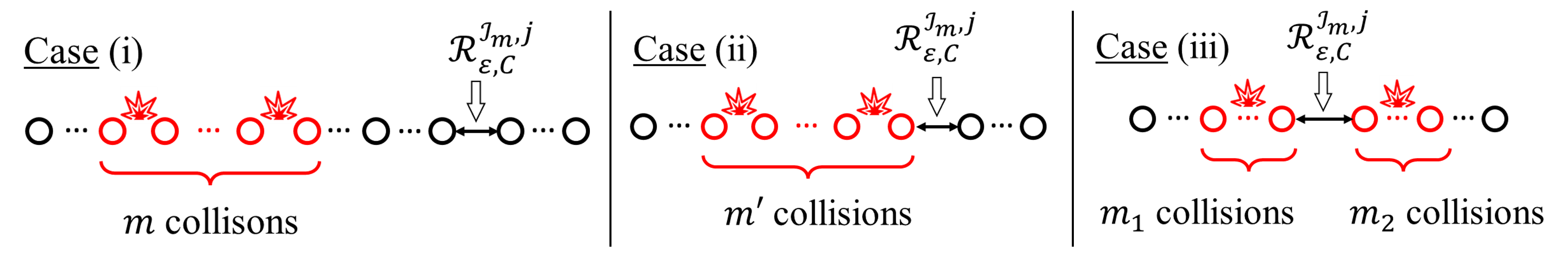}
	\caption{This figure depicts all three cases for the relative location between failure group(s) and the pair of the interest.}
	\label{fig:complete_cases}
\end{figure}

Observing the above result, the covariance matrix $\Sigma$ exhibits a distinctive symmetric tridiagonal structure, enabling the establishment of a closed-form expression for its inverse\footnote{In this section, we specifically address collisions as the existing failures. Nevertheless, one can extend the findings to encompass other events by following the similar analogous lines of reasoning.}. We now examine three possible scenarios regarding the relative position of the $j$'th pair in relation to the existing collisions $\mathcal{I}_m$: (i) The $j$'th pair is not adjacent to any of the collisions, (ii) The $j$'th pair is only adjacent to $m'$ consecutive collisions on one side, and (iii) The $j$'th pair is surrounded by $m_1$ and $m_2$ consecutive collisions. These three scenarios are visually depicted in Fig. \ref{fig:complete_cases}.

Let us construct a covariance matrix, denoted as $\hat{\Sigma}$, similar to the one defined in \eqref{eq:new-sig}. However, in this case, $\hat{\Sigma}$ only considers the adjacent collisions of size $\hat{m}$ and the $j$'th pair itself, which is a square matrix of size $(\hat{m}+1) \times (\hat{m}+1)$. The possible values for $\hat{m}$ are as follows: $\hat{m} = 0$ for Case (i), $\hat{m} = m'$ for Case (ii), and $\hat{m} = m_1+m_2$ for Case (iii). In all cases, we have $m \geq \hat{m}$. We denote the inverse of the submatrix $\hat{\Sigma}_{22}$ as $\hat{\Sigma}_{22}^{-1}$, which can be represented by the elements $[\hat{\alpha}_{ij}]$. The following result presents an analysis of the risk of cascading collision in a complete graph.

\begin{theorem}   \label{thm:mul_risk_complete}
    Suppose that the platoon \eqref{eq:dyn} reaches the steady-state with a complete communication graph and pairs $\mathcal{I}_m$ have experienced collisions $\bar{\bm{d}}_{\mathcal{I}_m} = \bm{0}_m$. The risk of cascading collision at the $j$'th pair is
    \begin{equation}
        \begin{aligned}
            \mathcal{A}_{\varepsilon}^{i,j} :=\begin{cases}
            0, &\text{if} ~ \frac{c\hat{\mu}-r}{c \hat{\sigma}} \geq \kappa_{\varepsilon}\\
            \dfrac{r}{\hat{\mu}-\kappa_\varepsilon \hat{\sigma}} - c, &\text{if} ~ \kappa_{\varepsilon} \in \big(\frac{c\hat{\mu} - r}{c \hat{\sigma}}, \frac{\hat{\mu}}{\hat{\sigma}} \big) \\
            \infty, &\text{if} ~ \frac{\hat{\mu}}{\hat{\sigma}} \leq  \kappa_{\varepsilon}
            \end{cases},
        \end{aligned}
    \end{equation}
    where $\kappa_{\varepsilon} = \left(\sqrt{2\pi} \varepsilon \exp(\io^2) \right)^{-1}$, $\io = \textup{erf}^{-1} (2\varepsilon-1)$, $\hat{\mu}$ and $\hat{\sigma}$ can be computed as follows: 
    
    \vspace{0.1cm}
    \noindent \underline{\textup{Case (i)}}: If $\hat{m} = 0$, then $\hat{\mu} = r$ and $\hat{\sigma} = \sqrt{\sigma_c}$. The risk of cascading collision remains unaffected.
    
    \vspace{0.1cm}
    \noindent\underline{\textup{Case (ii)}}: If $\hat{m} = m'$, i.e., there exists only one $k \in \mathcal{I}_m$ such that $|k-j| = 1$, one has
    \begin{equation}
        \hat{\sigma} = \sqrt{\sigj^2 - \frac{\sigma_c m'}{2(m'+1)}},
        ~
        \text{ and }
        ~ \hat{\mu} = r - \frac{\sigma_c}{2} \sum_{l=1}^{m'} \hat{\alpha}_{1,l} \, r.
    \end{equation}
    
    \vspace{0.1cm}
    \noindent \underline{\textup{Case (iii)}}: If $\hat{m} = m_1+m_2$, i.e., there exist $k = j - 1$ and $k' = j + 1$ for some $k,k' \in \mathcal{I}_m$, one has
    \begin{equation}
        \begin{aligned}
            \hat{\sigma} &= \sqrt{\sigi^2 - \frac{\sigma_c}{2} \frac{4m_1m_2 + m_1 + m_2}{m_1 + m_2 + 1}},\\
            \hat{\mu} &= r - \frac{\sigma_c}{2} \sum_{l=1}^{m_1+m_2} (\hat{\alpha}_{m_1,l} + \hat{\alpha}_{m_1+1,l} ) \, r.
        \end{aligned}
    \end{equation}
\end{theorem}

\begin{figure*}[t]
    \begin{subfigure}[t]{.32\linewidth}
        \centering
    	\includegraphics[width=\linewidth]{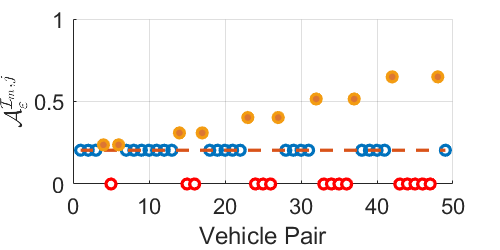}
    	\caption{Case (ii) with various values of $m'$.}
    \end{subfigure}
    \hfill
    \begin{subfigure}[t]{.32\linewidth}
        \centering
    	\includegraphics[width=\linewidth]{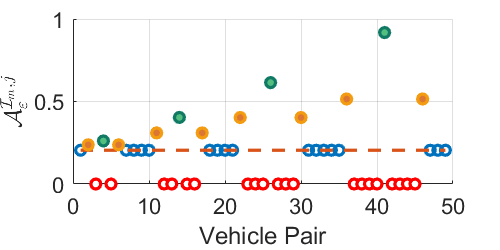}
    	\caption{Case (iii) with various values of $m_1 = m_2$.}
    \end{subfigure}
    \hfill
    \begin{subfigure}[t]{.32\linewidth}
        \centering
    	\includegraphics[width=\linewidth]{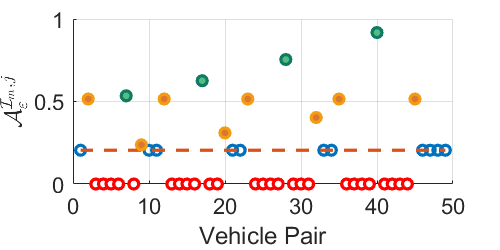}
    	\caption{Case (iii) with various values of $m_1 \neq m_2$.}
    \end{subfigure}
    \caption{Risk profile of cascading collisions with a complete communication graph.}
    \label{fig:mul_risk_complete}
\end{figure*}

The above result indicates that when the locations of existing collisions $\mathcal{I}_m$ are not adjacent to the $j$'th pair, the level of $\casm$ remains unchanged compared to the risk of single collision \cite{Somarakis2020b}. This is due to the vanishing cross-correlation in the complete graph when the pairs are not adjacent. This observation is supported by Lemma \ref{lem:sig_complete} when $|i - j| > 1$, as well as by Fig. \ref{fig:mul_risk_complete}.
In the second case, when the $j$'th pair is adjacent to only one specific "group" of collisions with size $m'$, the magnitude of $\casm$ is solely determined by the dimension of the collided group. This conclusion holds regardless of whether the $j$'th pair is positioned at the front or back of the collided group, see Fig. \ref{fig:mul_risk_complete} (a).
In the last case, the $j$'th pair is situated between two distinct clusters of collisions, with sizes $m_1$ and $m_2$ respectively. The combined impact of these two collided groups contributes to the risk of cascading collisions, which, as illustrated by the green dots in Fig. \ref{fig:mul_risk_complete} (b), varies based on the values of $m_1$ and $m_2$.

\subsection{The Path Graph}
If a platoon of $n$ vehicles are using an unweighted path graph for communication, which contains $n$ nodes and $n-1$ edges. The eigenvalues are: $\lambda_j = 2(1-\cos(\pi(j-1)/n))$, and their corresponding eigenvectors are shown by $\bm{q}_1 = \frac{1}{\sqrt{n}} \bm{1}_n$ and $\bm{q}_k = [q_k^{(1)}, \dots, q_k^{(n)}]^T$ with $q_k^{(l)} = \sqrt{\frac{2}{n}} \cos \left( \frac{\pi(n-j+1)}{2n} (2l-1) \right)$ for all $k = 2, \dots, n$ and $l = 1, \dots, n$. For the steady-state inter-vehicle distance, the elements of the covariance matrix are
\begin{equation*}
        \sigma_{ij} = \frac{4 g^2 \tau^{3}}{n\pi} \sum_{k=2}^{n} \theta(i,j,k)  f(\lambda_k \tau, \beta \tau),
\end{equation*}
in which $\theta(i,j,k) = \sin (\frac{\pi(n-k+1)}{n} i) \sin (\frac{\pi(n-k+1)}{n} j)$ $\sin^2 (\frac{\pi(n-k+1)}{2n})$. When the collision happens at the $i$'th pair, the conditional distribution on the $j$'th pair is given by
\begin{align*}
    \hat{\mu} = \left(1 - \frac{\sum_{k=2}^n \theta(i,j,k)f(\lambda_k \tau, \beta \tau)}{\sum_{k=2}^n \theta(i,i,k)f(\lambda_k \tau, \beta \tau)} \right) \, r,
\end{align*}
and 
\begin{equation*}
    \begin{aligned}
        \hat{\sigma}^2 \! = \! \frac{4 g^2 \tau^{3}}{n\pi} \! \! \left( \sum_{k=2}^{n} \theta(j,j,k)  f(\lambda_k \tau, \beta \tau) \! - \! \frac{(\sum_{k=2}^n \theta(i,j,k)f(\lambda_k \tau, \beta \tau))^2}{\sum_{k=2}^n \theta(i,i,k)f(\lambda_k \tau, \beta \tau)} \! \right) \! \!.
    \end{aligned}
\end{equation*}
The risk of cascading collisions can be evaluated using Theorem \ref{thm:avar_single_cas} or Corollary \ref{cor:risk_mul_cas}, and some of examples are shown in both Example 1 and 2.

\subsection{The p-Cycle Graph}
If a platoon of vehicles are using an unweighted $p-$cycle graph for communication, which denotes each vehicle communicates to its $p-$immediate neighbors. From \cite{gray2006toeplitz, van2010graph}, one can show that eigenvalues of the corresponding Laplacian matrix are $\lambda_1 = 0$ and $\lambda_k = \big(\frac{\sin((2p+1)(k-1)\pi/n)}{\sin((k-1)\pi/n)} - 1\big)$. Their corresponding eigenvectors are shown by $\bm{q}_1 = \frac{1}{\sqrt{n}} \bm{1}_n$ and $\bm{q}_k = [q_k^{(1)}, \dots, q_k^{(n)}]^T$ with $q_k^{(l)} = \big(2p+1 - \frac{\sin((2p+1)(l-1)\pi/n)}{\sin((l-1)\pi/n)} \big)$ for all $k \in \{2, \dots, n\}$ and $l \in \{1, \dots, n\}$. The elements of the covariance matrix $\Sigma$ is shown by
\begin{equation*}
    \sigma_{ij} = \frac{g^2 \tau^{3}}{2\pi} \sum_{k=2}^{n} \big(q_k^{(i+1)} - q_k^{(i)} \big) \big( q_k^{(j+1)} - q_k^{(j)} \big)  f(\lambda_k \tau, \beta \tau).
\end{equation*}
When the $i$'th pair has collided, the conditional statistics of the $j$'th pair is given by
\begin{equation*}
    \begin{aligned}
        \hat{\mu} = \left(1-\frac{\sum_{k=2}^{n} \big(q_k^{(i+1)} - q_k^{(i)} \big) \big( q_k^{(j+1)} - q_k^{(j)} \big)  f(\lambda_k \tau, \beta \tau)}{\sum_{k=2}^{n} \big(q_k^{(i+1)} - q_k^{(i)} \big)^2  f(\lambda_k \tau, \beta \tau)} \right)r,
    \end{aligned}
\end{equation*}
and 
\begin{equation*}
    \begin{aligned}
        \hat{\sigma}^2 = \frac{g^2 \tau^3}{2 \pi}  \bigg( \sum_{k=2}^{n} (q_k^{(j+1)} - q_k^{(j)})^2  f(\lambda_k \tau, \beta \tau) - \frac{(\sum_{k=2}^{n} (q_k^{(i+1)} - q_k^{(i)}) (q_k^{(j+1)} - q_k^{(j)}) f(\lambda_k \tau, \beta \tau))^2}{\sum_{k=2}^{n} (q_k^{(i+1)} - q_k^{(i)})^2  f(\lambda_k \tau, \beta \tau)} \bigg).
    \end{aligned}
\end{equation*}
Then, one can compute the risk of cascading collisions as in Theorem \ref{thm:avar_single_cas} or Corollary \ref{cor:risk_mul_cas}.

\begin{figure}[t]
    \centering
    \includegraphics[width=0.75\linewidth]{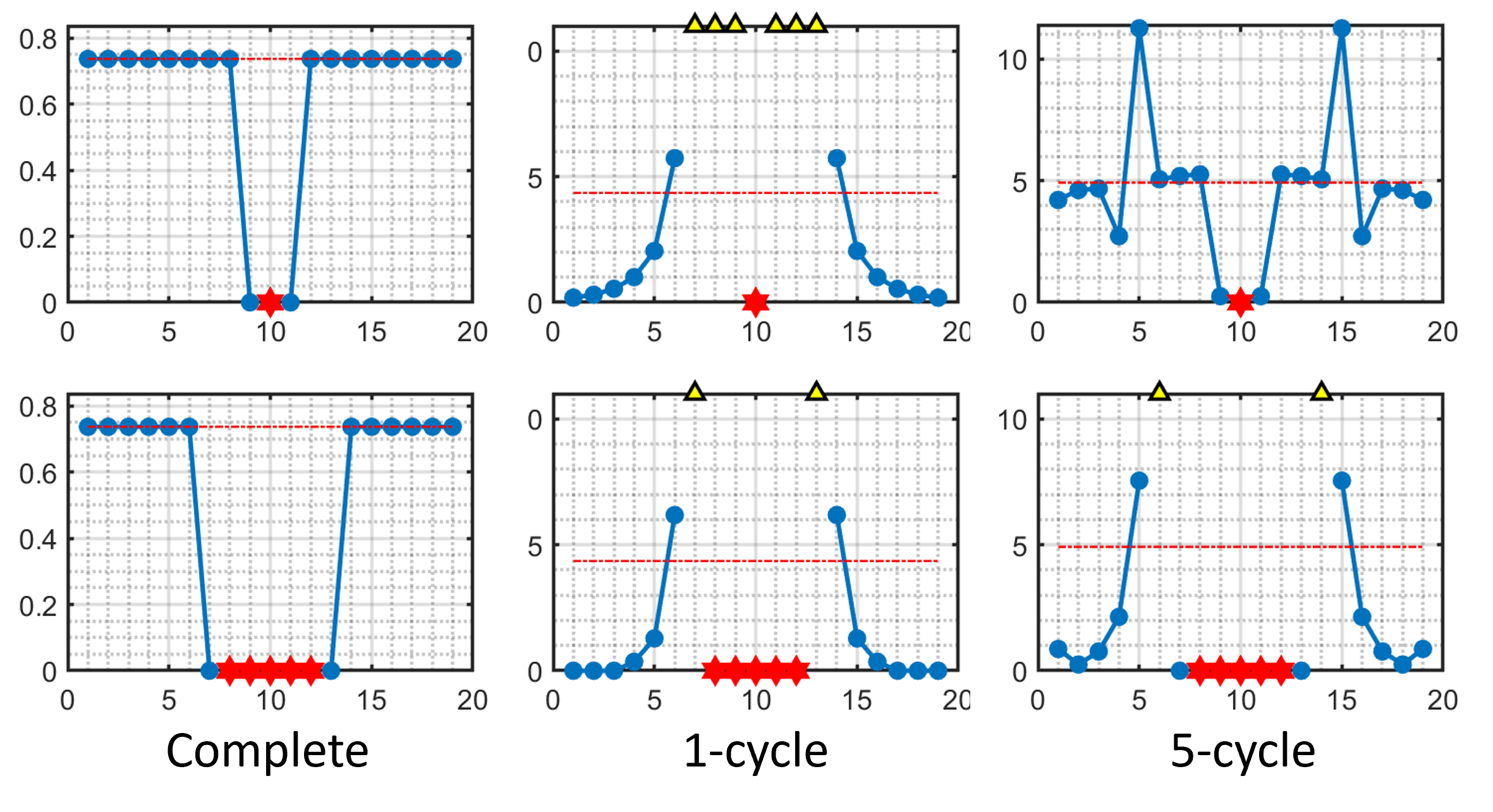}
    \caption{{ The risk of cascading collisions $\cass$ (top) and $\casm$ (bottom, assuming failures occur at pairs $\{8,9,10,11,12\}$). The red dash represents the risk of collision without prior failures.}}
    \label{fig:risk_collision}
\end{figure}

\vspace{2mm}

\noindent {\bf Example 3:}
Some examples of using the above communication graphs are simulated with $n=20, c = 1.1, r = 2, \tau = 0.04, \beta = 1, \text{ and } \varepsilon = 0.1$, as shown in Fig. \ref{fig:risk_collision}. We evaluate the risk profile of cascading collision $\cass$ assuming that the middle pair of vehicles has collided. For multiple observed pairs, we examine vehicle pairs labeled as $\mathcal{I}_m = \{8,9,10,11,12\}$ have collided.

\noindent \underline{Complete Graph}: 
In a complete graph, we set $g = g_1 = ... = g_n = 10$. It is shown that the immediate neighbor of an existing collision, denoted as $\cass$, exhibits a lower risk compared to that of a single collision, denoted as $\srisk$. Furthermore, the risk of cascading collisions on the remaining pairs remains unchanged when compared to $\srisk$. When the platoon experiences multiple collisions, the adjacent vehicle pairs to the failure group demonstrates a lower risk. Furthermore, the remaining pairs share the same value as $\srisk$, which is evident in Theorem \ref{thm:mul_risk_complete}. 

\noindent \underline{p-Cycle Graph}:
We set $g=1$ for a $1-$cycle graph and { $g_1 = ...= g_n = 10$} for a $5-$cycle graph. This type of communication resemble a platoon formation model where vehicles are arranged in a loop and have the ability to communicate with their $p$ immediate neighbors \cite{wu2017flow}. The obtained results exhibit certain properties resembling those of a complete graph. Specifically, the immediate neighbor of the collision experiences a lower $\cass$ compared to the risk of single collision. With smaller $p$ values, vehicles face higher risks when they are in proximity to existing collisions. Furthermore, an interesting observation is that as the value of $p$ increases, the risk profile exhibits a pattern more akin to that of a complete graph.

\section{Time-delay induced Fundamental Limits on the Risk of Cascading Collisions}     \label{sec:limits}

From an engineering standpoint, one does not have the control over the communication time-delay and external disturbances. Hence, when designing the network, the focus shifts towards modifying communication topologies to achieve optimal performance and safety. In this study, our objective is to investigate the minimum achievable risk of cascading collisions for general communication topologies with { $g = g_1 = ... = g_n$}. We will first establish our result by initially considering the general graph topologies and then present a concrete example using the complete communication graph.

\subsection{Fundamental Limits on the Steady-state Covariance}

In this subsection, we study the boundedness of the steady-state covariance of the inter-vehicle distances on a { convex} and compact subset of the stability set $S$. We recall that $f(s_1,s_2)$ diverges on the boundary of $S$ for $s_2 \neq 0$ where $f$ meets its poles. On the $s_1$ axis, over which $s_2 = 0$ and $f$ is finite, the dynamic of the unperturbed platoon will fail to reach the steady-state. Thus, we will consider a compact subset of $S$ as 
\begin{equation} 
    \bar{S} = [0.1, s^*\sin(s^*)-0.1] \times [0.1,0.9],
\end{equation}
where $s^*$ is the solution of $s^* \cot(s^*) = s_2$ for $s_2 \in [0.1,0.9]$. One can verify that $s^* \cot(s^*) = s_2$ is invertible for $s_2 \in [0.1, 0.9]$ and can be written as $s^*(s_2)$. Then, we consider the family of all connected communication graphs such that the following condition is satisfied 
\begin{align}   \label{eq:bar_stable}
    \Big(\lambda_i \, \tau, \beta \, \tau \Big) \in \bar{S},
\end{align}
for all $i=2,...,n$.

In the following result, we will consider all the communication graphs that satisfy $(s_1,s_2) \in \bar{S}$, and reveal the time-delay induced fundamental limits on those graphs. The following result explains time-delay induced limits on the elements of the inter-vehicles covariance matrix.

\begin{theorem} \label{thm:sig_upper_lower_limit}
For a platoon \eqref{eq:dyn} with any connected communication graph that reaches the steady-state, every $\sigma_{ij}$ of the steady-state covariance matrix $\Sigma$ satisfies the following limits:
\begin{equation}
    \left\{
        \begin{aligned}
        \underline{\sigma}-\bar{\sigma} \leq &~\sigma_{ij} \leq \bar{\sigma} - \underline{\sigma}, &\text{if} ~ |i-j| > 1\\[5pt]
        \frac{1}{2}\underline{\sigma} - \frac{3}{2}\bar{\sigma} \leq &~\sigma_{ij} \leq \frac{1}{2}\bar{\sigma} - \frac{3}{2}\underline{\sigma},  &\text{if} ~ |i-j| = 1\\[5pt]
        2\underline{\sigma} \leq &~\sigma_{ij} \leq 2\bar{\sigma}, &\text{if} ~ |i-j| = 0
    \end{aligned}  \right.
\end{equation}
    where $\underline{\sigma} = g^2\frac{\tau^3}{2\pi} \underline{f},$ and $\bar{\sigma} = g^2\frac{\tau^3}{2\pi} \bar{f}.$ 
    The value $\underline{f}:= \inf_{(s_1,s_2)\in \bar{S}} f(s_1,s_2)$ $ \approx 25.4603$ is a lower bound of $f$, and $\bar{f}:= \sup_{(s_1,s_2)\in \bar{S}} f(s_1,s_2)$ $ \approx 3.633\times10^3$ is the upper bound of $f$. 
\end{theorem}

The above theorem establishes both upper and lower limits for the entries of the covariance matrix. It is noteworthy that these limits depend on factors such as the relative position of vehicles within the communication graph $|i-j|$, the time-delay $\tau$, and the noise magnitude $g$. We highlight that the above result holds true for any connect graphs as long as it satisfies the stability condition \eqref{eq:bar_stable}.

\subsection{A Lower Estimate of the Best Achievable Risk of Cascading Collision}
Considering the time-delay and the graph connectivity induced limits on the covariance of the steady-state inter-vehicle distances, there exists limitation on the best achievable risk of cascading collisions among all feasible communication graph designs. In the following result, we provide a convenient lower estimate of the best achievable risk of cascading collision by considering the scenario featuring a single existing collision, as formulating the conditional distribution with multiple existing events introduces challenges in tracing the boundedness of the covariance terms.

\begin{theorem}     \label{thm:cas_bound}
    Suppose that the platoon \eqref{eq:dyn} reaches the steady-state and the $i$'th pair has collided, i.e., $\bar{d}_i = 0$. Then, a lower estimate the best achievable risk of cascading collision at the $j$'th is characterized as follows: 
    
    \noindent If $\sigma_{ij} > 0$,
    \begin{equation}
        \begin{aligned}
            \mathcal{A}_{+} := \begin{cases}
            0, &\text{if} ~ 1-\frac{\sqrt{\underline{\sigma}}}{\sqrt{\bar{\sigma}}} \geq \frac{1}{c}\\
            \dfrac{\sqrt{\bar{\sigma}}}{\sqrt{\bar{\sigma}} - \sqrt{\underline{\sigma}}} - c, &\text{if} ~ 1-\frac{\sqrt{\underline{\sigma}}}{\sqrt{\bar{\sigma}}} \in \big(0, \frac{1}{c}\big) 
            \end{cases}.
        \end{aligned}
    \end{equation}

    \noindent If $\sigma_{ij} < 0$,
    \begin{equation}
        \mathcal{A}_{-} := 0.
    \end{equation}
    
    \noindent If $\sigma_{ij} = 0$,
    \[
        \mathcal{A}_{0} := \begin{cases}
            0, &\text{if} ~ \frac{cr-r}{c\sqrt{\underline{2\sigma}}} \geq \kappa_\varepsilon\\
            \dfrac{r}{r - \kappa_\varepsilon \sqrt{2 \underline{\sigma}}} - c, &\text{if} ~ \kappa_\varepsilon \in \left(\frac{cr-r}{c\sqrt{2\underline{\sigma}}}, \frac{r}{\sqrt{2\underline{\sigma}}} \right) \\
            \infty,&\text{if} ~ \frac{r}{\sqrt{\underline{\sigma}}} \leq \kappa_\varepsilon
            \end{cases}.
    \]
    The values of $\bar{\sigma}$ and $\underline{\sigma}$ are as in Theorem \ref{thm:sig_upper_lower_limit}.
\end{theorem}


The above result presents a lower estimate of the best achievable risk of cascading collision for any communication graphs that satisfies the stability condition \eqref{eq:bar_stable}, indicating that the risk of cascading collision can not be further optimized beyond this threshold by altering the communication graph design. Unlike the previous analyzed time-delay induced fundamental limits on the risk of single collision \cite{Somarakis2020b}, the best achievable risk of cascading collision does not obtain a uniform value since the sign of the cross-correlation between vehicle pairs alters. This implies that when $\mathcal{A}_{\varepsilon}^{i,j}$ reaches its lower limit, e.g., $0$, it does not indicate that the risk of cascading failure on the other vehicle will reach the same limit. 

Considering the fact that the above result provides a lower estimate instead of the actual best achievable risk of cascading collision, one can not guarantee if there exists a communication graph that can reach the lower estimate. However, when designing a platoon \eqref{eq:dyn} with certain constraints on the risk of cascading collision, one may consider 
\[
    \max \{\mathcal{A}_{+}, \mathcal{A}_{0}, \mathcal{A}_{-}\}
\]
to rule out the infeasible risk-aware design target without validating over all potential communication graphs. By imposing certain symmetry assumption to the communication graph, one can narrow down the boundedness presented in Theorem \ref{thm:sig_upper_lower_limit} and specify the best achievable risk of cascading collision in Theorem \ref{thm:cas_bound}, as discussed in the subsequent section.
 
\subsection{Best Achievable Risk of Cascading Collision on the Complete Graph}

When the underlying graph is an unweighted complete graph, one can obtain the explicit formulas of $\sigma_{ij}$ using Lemma \ref{lem:sig_complete}. Then, the best achievable risk of cascading collisions can be further specified, as presented in the following result. 
 
\begin{corollary}     \label{cor:limit_complete}
Suppose that the platoon \eqref{eq:dyn} reaches the steady-state with a complete communication graph and the $i$'th pair is observed with $\bar{d}_i = d^*$. Then, the best achievable risk of cascading collision on the $j$'th pair can be characterized as follows: 
\vspace{0.1cm}

\noindent{\underline{\textup{Case(i):}}} If $|i-j|> 1$,
    \[
        \cass \geq \begin{cases}
            0, &\text{ if} ~ \frac{cr-r}{c\sqrt{2\underline{\sigma}}} \geq \kappa_\varepsilon \\
            \frac{r}{r - \kappa_\varepsilon \sqrt{2\underline{\sigma}}} -c, &\text{ if} ~ \kappa_\varepsilon \in \left(\frac{cr-r}{c\sqrt{2\underline{\sigma}}}, \frac{r}{\sqrt{2\underline{\sigma}}} \right)\\
            \infty, &\text{ if} ~ \frac{r}{\sqrt{2\underline{\sigma}}} \leq \kappa_\varepsilon
            \end{cases}.
    \]
\vspace{0.1cm}
    
\noindent{\underline{\textup{Case(ii):}}} If $|i-j| = 1$,
    \begin{equation*}
        \cass \geq \begin{cases}
        0, &\text{ if} ~ \frac{c(3r-d^*)-2r}{c\sqrt{6\underline{\sigma}}} \geq \kappa_\varepsilon\\
        \frac{2r}{3r-d^*-\kappa_\varepsilon \sqrt{6 \underline{\sigma}}} -c, &\text{ if} ~ \kappa_\varepsilon \in \left(\frac{c(3r-d^*)-2r}{c\sqrt{6\underline{\sigma}}}, \frac{3r-d^*}{\sqrt{6 \underline{\sigma}}} \right)\\
        \infty, &\text{ if} ~ \frac{3r-d^*}{\sqrt{6 \underline{\sigma}}} \leq \kappa_\varepsilon
        \end{cases},
    \end{equation*}
    where $\underline{\sigma}$ is as in Theorem \ref{thm:sig_upper_lower_limit}.
\end{corollary}

The above result characterizes the best achievable risk of cascading collisions when employing a complete graph as the communication topology, while taking into account the communication time-delay and external disturbance. When $|i-j|>1$, one can observe that the risk expression which boils down to the best achievable risk of single collision \cite{Somarakis2020b}, or the case when $\sigma_{ij} = 0$ in Theorem \ref{thm:cas_bound}, since the distribution of $i$'th and the $j$'th vehicle are not correlated in a complete communication graph. For the case when $|i-j| = 1$, the best achievable risk can be obtained when $\sigma_c = 2\underline{\sigma}$, and such limit is uniform over among the entire platoon.

\section{Case Studies}  \label{sec:case}

{ We discuss the case studies for platoons with dynamics governed by \eqref{eq:dyn} among the various communication graphs with $n=20, c = 1.1, r = 2, \tau = 0.04, \beta = 1, g= g_1 = ...=g_n \text{ and } \varepsilon = 0.1$.}

\begin{figure*}[t]
    \centering
    \includegraphics[width=\linewidth]{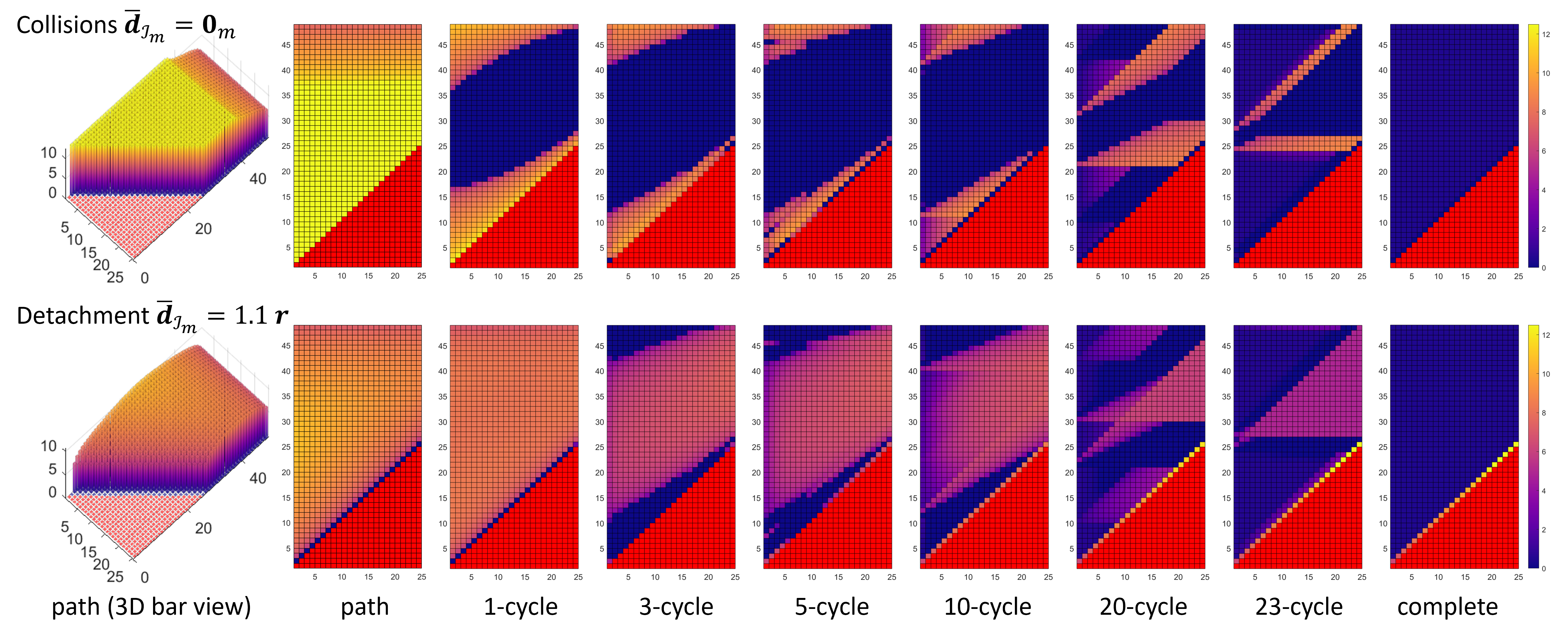}
    \caption{{ Risk profile among various scale of failures and communication graphs with the red region denotes the failed vehicle pairs. The top tow resembles the risk when existing failures are collisions $(\bm{0})$, and the bottom row represents the inter-vehicle detachment $(1.1r \bm{1}_m)$ \cite{Somarakis2020b}.}}
    \label{fig:num_failures}
\end{figure*}

\begin{figure}[t]
    \centering
    \includegraphics[width=\linewidth]{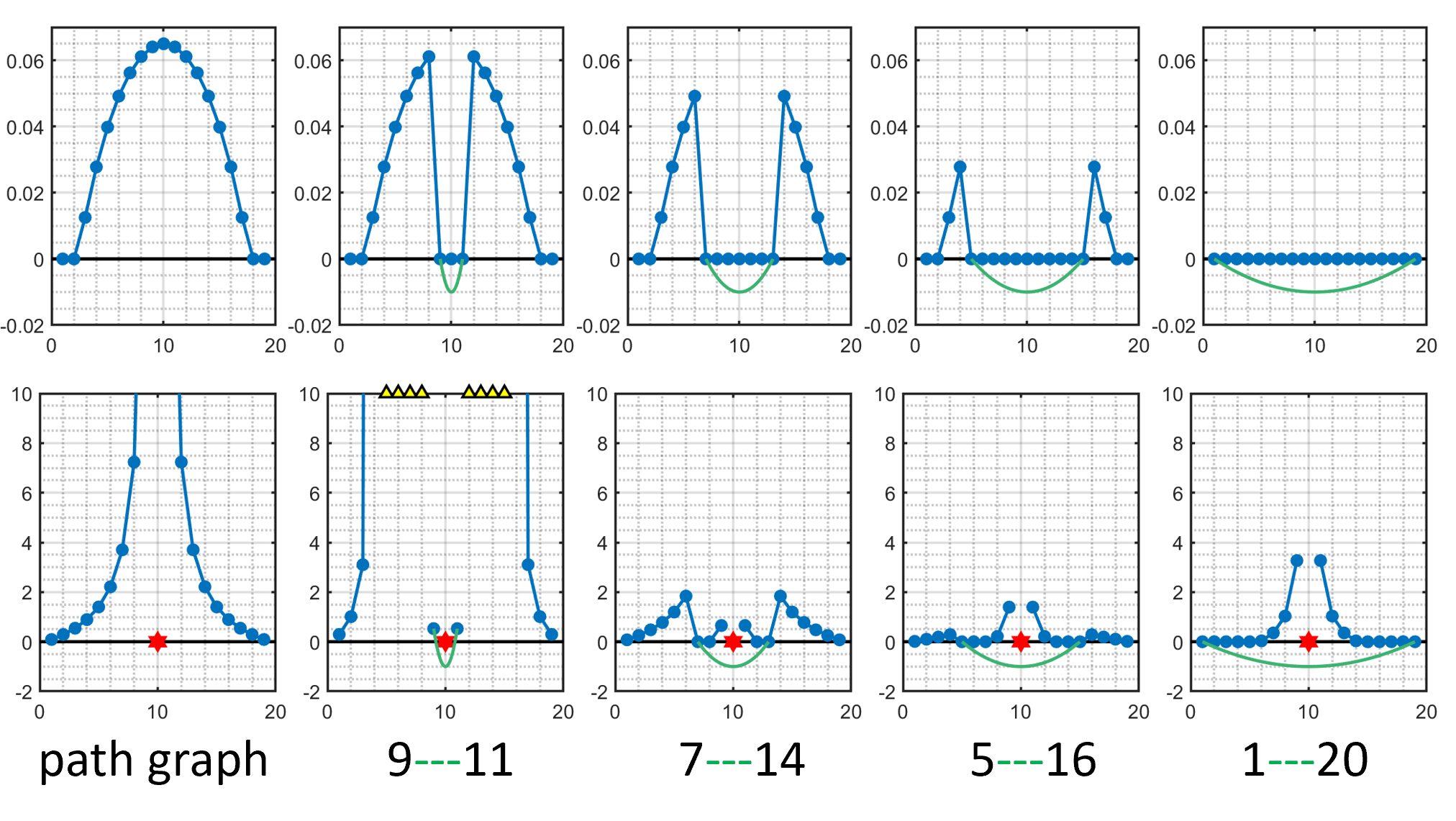}
    \caption{{ Risk profile when adding edges (dark green solid line) to the path graph. The top row represents the risk of single collision, and the bottom row represents the risk of the cascading collision.}}
    \label{fig:add_edge}
\end{figure}

\begin{figure}[t]
    \centering
    \includegraphics[width=\linewidth]{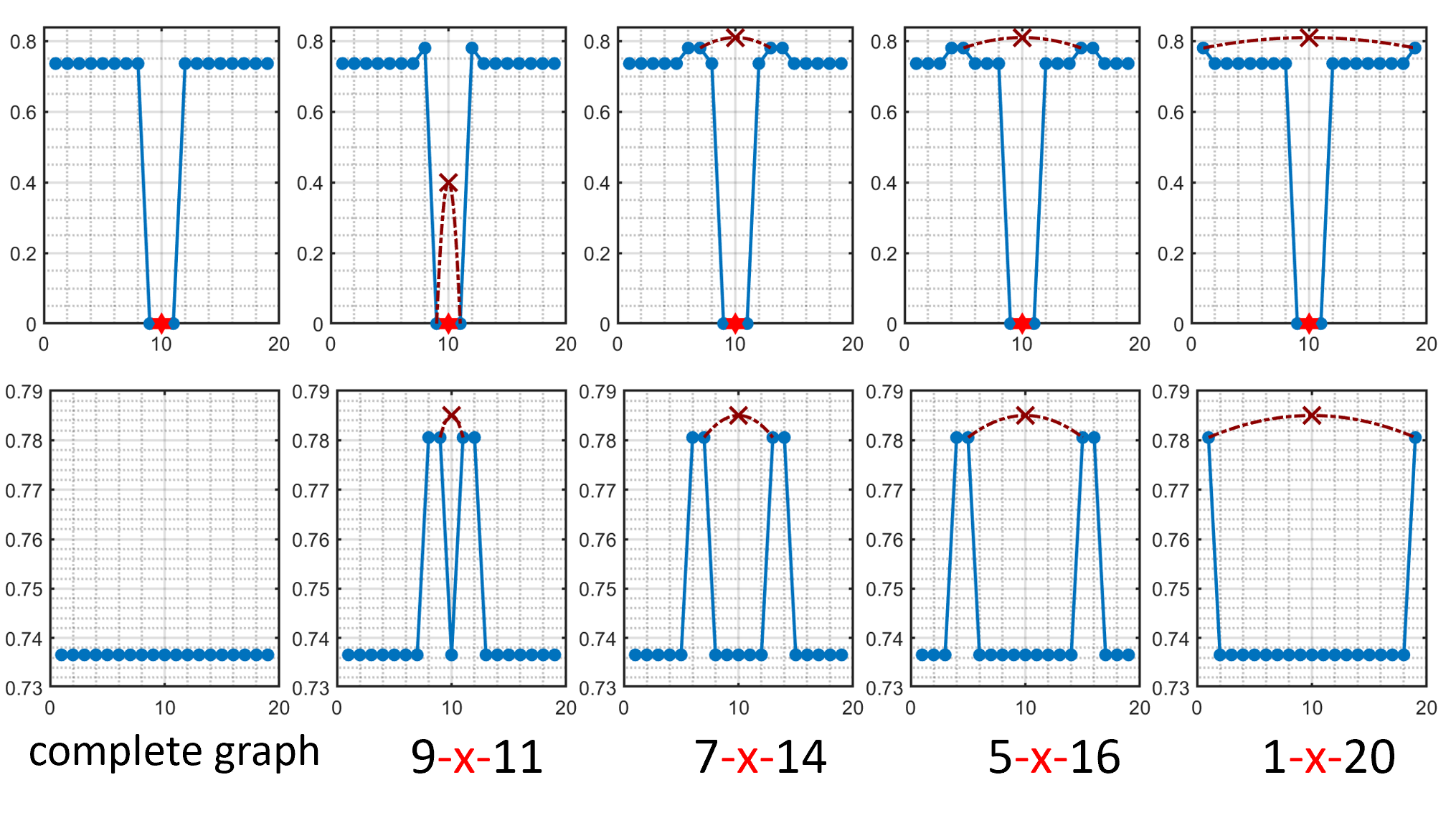}
    \caption{{ Risk profile when removing edges (dark red dashed line) from the complete graph. The top row represents the risk of single collision, and the bottom row represents the risk of the cascading collision.}}
    \label{fig:remove_edge}
\end{figure}

\begin{figure*}[t]
    \centering
    \includegraphics[width=\linewidth]{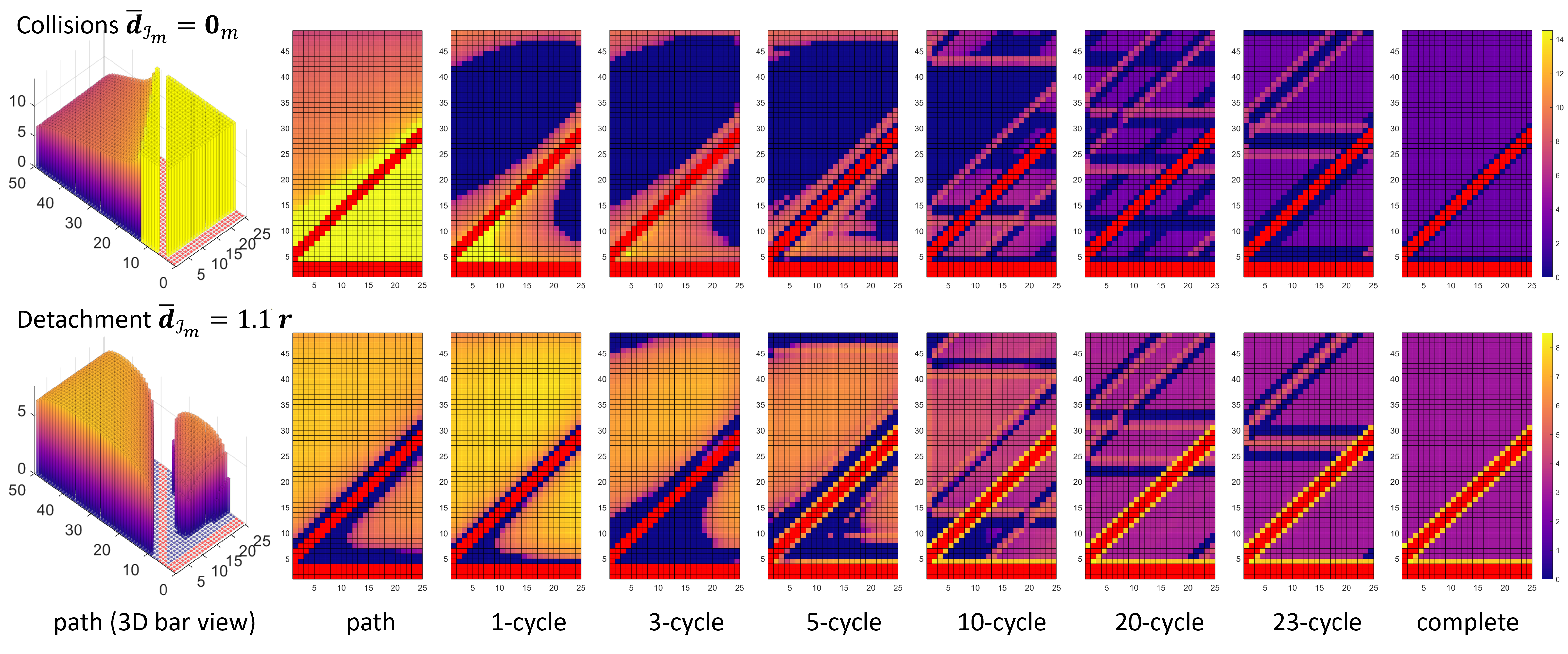}
    \caption{{ Risk profile with various spatial distribution of the existing failures. The red region denotes the vehicle pairs observed with failures.}}
    \label{fig:loc_failures}
\end{figure*}

\subsection{Impact from Failures' Proprieties}

In the case only one pair of observed vehicles, the impact of the observed vehicle pair to the platoon is limited since it can only vary in either the position within the communication network or the value of the observed inter-vehicle distance. However, the impact resulting from malfunctioning vehicles becomes increasingly complex when the dimension of the failure extends beyond a single pair to {\it an arbitrary} number, denoted as $m < n$. One can anticipate the cascading effect is influenced by the state of the observed vehicle pairs, the spatial distribution of the affected vehicles, and the underlying communication graph topology.

\subsubsection{Dimension of Observed Vehicle Pairs}

The dimension of the measurement of observed vehicle pairs significantly impacts $\casm$ due to its strong correlation with the overall uncertainty of the platoon and the conditional distribution of $\bar{d}_j$, both of which are functions of the value of $m$. In order to demonstrate this phenomenon, we consider the presence of inter-vehicle collisions ranging from $1$ to $10$ within the platoon. Furthermore, we analyze the risk profile across different communication graph topologies, as illustrated in Fig. \ref{fig:num_failures} (a) - (e). Several noteworthy observations emerge from this analysis. Firstly, in a path graph, the overall risk escalates as the magnitude of existing failures increases. Secondly, in a complete graph, the risk of collision for the adjacent pair to the failure group $\mathcal{I}_m$ diminishes, which aligns with the findings of Theorem \ref{thm:mul_risk_complete}. Thirdly, in the case of communication through a $p$-cycle graph, the pair situated in the middle of the functioning vehicles exhibits the highest level of safety.

\subsubsection{Spatial Distribution of the Existing Collisions}

When the scale of existing collisions is fixed for the system, the distribution of these collisions' locations significantly affects the risk profile. To assess this impact, we employ the following approach: assuming the occurrence of $m = 10$ collisions at the beginning of the platoon, and divide them into two clusters, each comprising $5$ failures. Subsequently, we gradually increase the distances between these clusters, as illustrated in Fig. \ref{fig:loc_failures} (a) - (e). In a complete graph, it is crucial to prioritize the non-adjacent pair to $\mathcal{I}_m$ since they obtains a higher risk of encountering the cascading collision. Alternatively, when employing a $p-$cycle graph, attention should be given to the pair that is close to the existing collisions. By adopting a path graph communication topology, the overall risk profile increases as the grouped failures become more spatially distributed within the platoon. Consequently, in applications involving highway platoons, it is advisable to prevent distributed collisions and rearrange the vulnerable pairs in the graph in a clustered manner.

\subsubsection{States of Existing Failures}

In addition to our previous work \cite{liu2022emergence,liu2021risk}, the proposed framework expands the scope beyond considering collisions exist in the platoon, allowing for the evaluation of risk of cascading failures in various scenarios. As stated in Sec. \ref{sec:risk}, the risk of cascading collisions can be evaluated with snapshots of vehicle pairs of the platoon, despite the value of the measured distances. To illustrate this concept, we simulate $\casm$ with $\bm{d}^* = 1.1 r \bm{1}_{m}$, representing situations where pairs are widely separated rather than colliding \cite{Somarakis2020b}, see Fig. \ref{fig:num_failures} (f) - (j) and Fig. \ref{fig:loc_failures} (f) - (j). Notably, the risk profile exhibits distinct features compared to scenarios involving existing events caused by collisions. For instance, in a complete communication graph, the vehicle pairs that are close to the detached vehicles experience a higher risk of cascading collision, which is contrary to the case when some vehicles are observed colliding to each other. In a path communication graph, i.e., a highway driving scenario, our obtained results indicate that the detached pairs will reduce the risk of vehicles that are close to them, and a more spatially distributed detached pairs will reduce the overall risk of cascading collisions for the platoon.

\subsection{Cascading Risk-aware Communication Graph Design}
In most real-world applications of vehicle platooning, the resources of reconstructing the communication graphs are limit to only adding or removing certain links. In this section, our results provide valuable insights on how adding or removing links on standard communication graphs will impact both the risk of collision or cascading collisions in a platoon.

\underline{Adding Links}:
In Fig. \ref{fig:add_edge}, we use path graph for example with $n=20, g=0.01, \beta =1, \text{ and } \tau = 0.04$. Let us first evaluate how the risk of cascading collision might be changed when edges are added to the graph. One can observe that enhancing the local communication will not always benefit the mitigation of the risk of cascading collision. On the other hand, the long range communication is not always the optimal solution for the path graph, indicating that the middle ranged communication is the optimal solution in this case. However, the risk of single collision obtains a different property, since increasing the length of the added communication link will always help to mitigate the risk of the single collision. Hence, it is immediate to reveal the potential trade-off between the risk of single and cascading collisions, i.e., the optimal risk profile are not obtained with the same communication graph structure for both risks and it is necessary to find the balance during the phase of communication graph design.

\underline{Removing Links}:
The effect of removing links is illustrated in Fig. \ref{fig:remove_edge}, in which the complete graph is adopted for communication with $n=20, g=10, \beta =1, \text{ and } \tau = 0.04$. In the case of single collision, the complete communication graph shows an invariant property as removing edges with any length will only affect the vehicle that was connected to it, and the magnitude of the risk increment remains the same. The same invariance also appears in the risk of cascading collisions in the complete graph, despite the length of the removed link, the increment of the risk of cascading collisions remains the same, and the rest of vehicles remains unchanged. Our finding advises that the complete communication graph is preferable in terms of robustness, as it is able to isolate and localize the impact of losing communication links.

\subsection{Non-uniform Input Noise Magnitude}
When considering the additive noise is given by $G = \text{diag} \left\{ g_{1}, ..., g_{n}\right\}$ as in \eqref{eq:dyn}. We observe the impact of various constructions of $G$ to the risk of cascading failures. Some examples of using common communication graphs are simulated with $n=20, c = 1.1, r = 2, \tau = 0.04, \beta = 1, \text{ and } \varepsilon = 0.1$. We evaluate the risk profile of cascading collision $\cass$ assuming that the middle pair of vehicles has collided. For multiple observed pairs, we examine vehicle pairs labeled as $\mathcal{I}_m = \{9,10,11\}$ have collided.

\begin{figure*}[t]
    \centering
    \includegraphics[width=\linewidth]{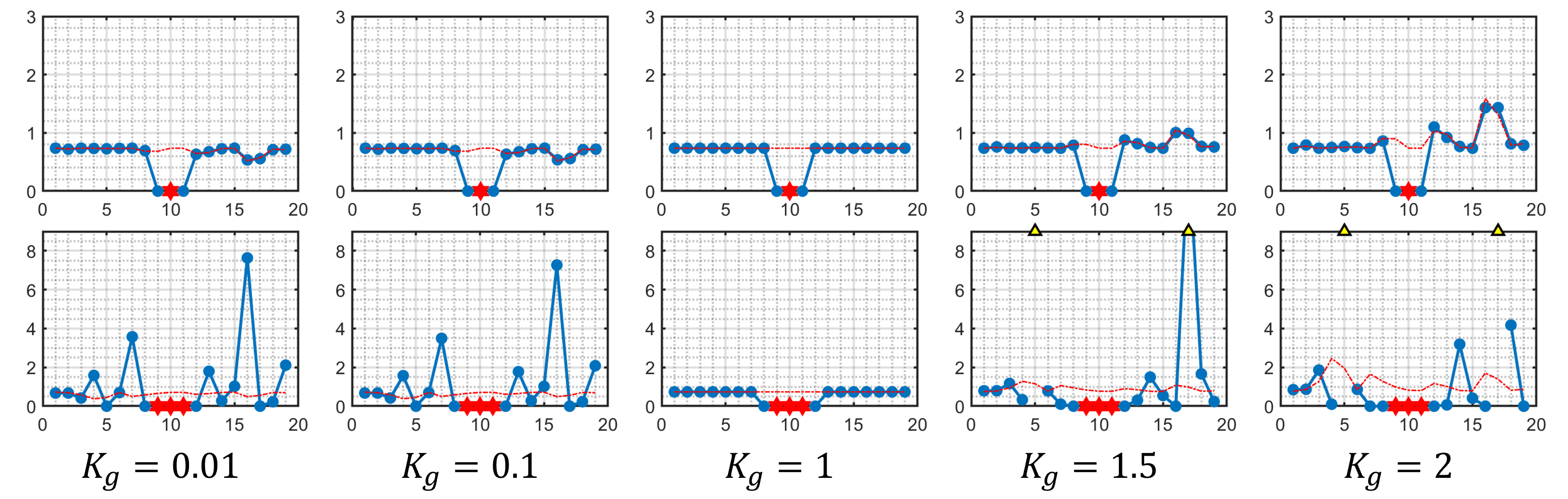}
    \caption{{ Risk profile with various $K_g$ of failing vehicles in a complete graph.}}
    \label{fig:complete_G}
\end{figure*}
\begin{figure*}[t]
    \centering
    \includegraphics[width=\linewidth]{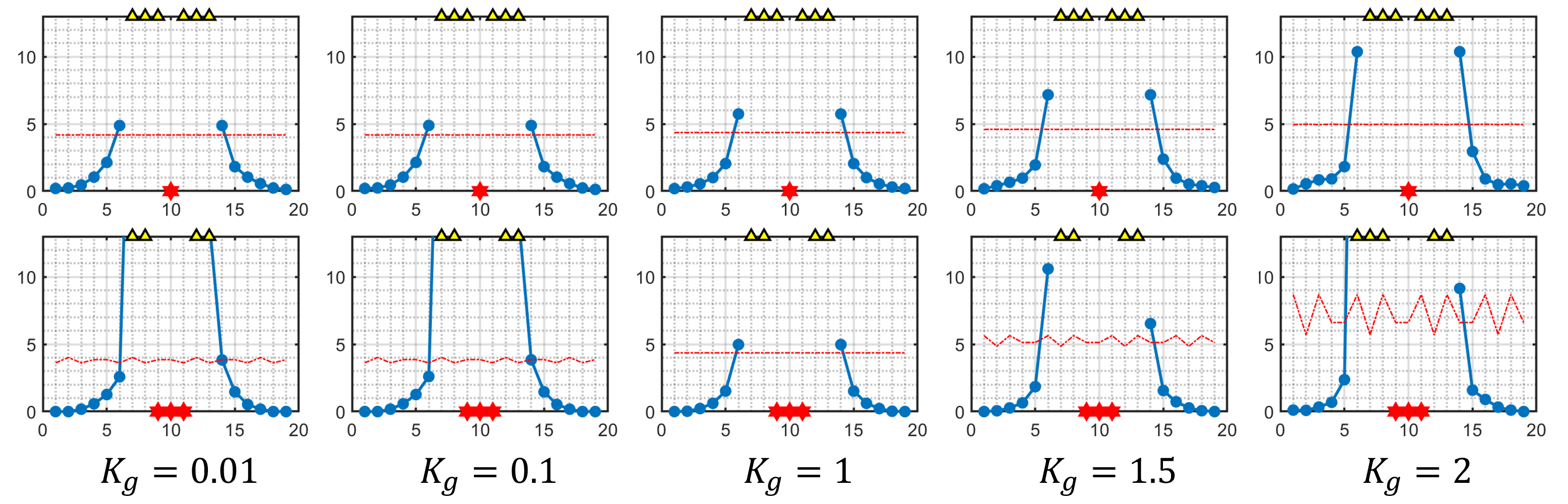}
    \caption{{ Risk profile with various $K_g$ of failing vehicles in a 1-cycle graph.}}
    \label{fig:cycle_G}
\end{figure*}
\begin{figure*}[t]
    \centering
    \includegraphics[width=\linewidth]{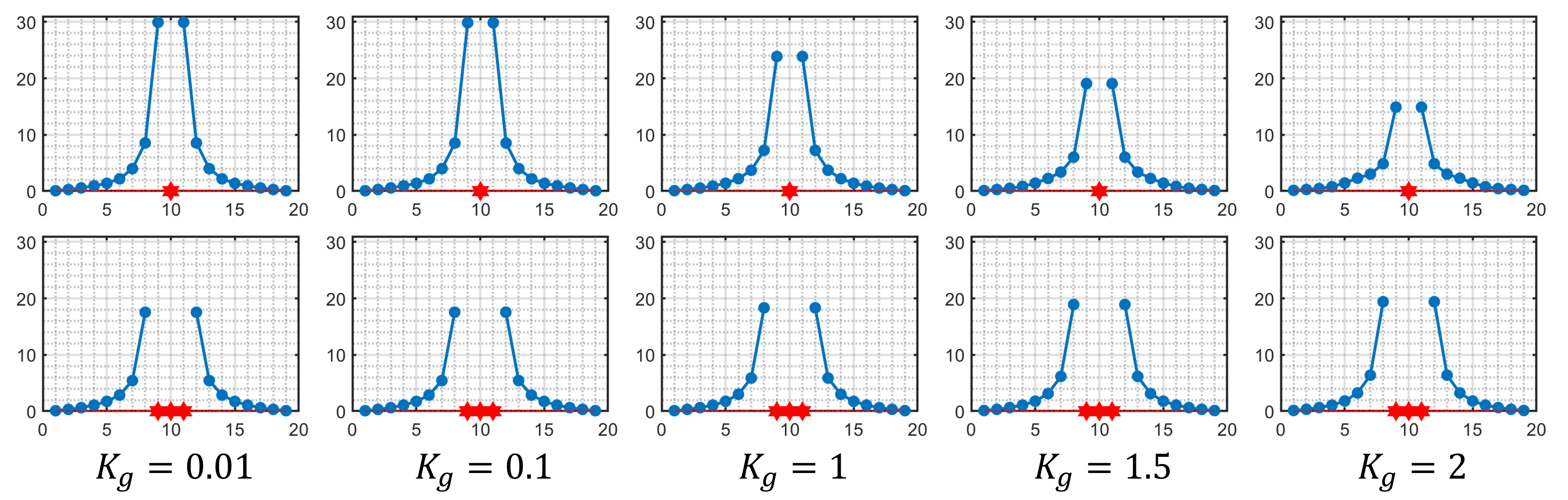}
    \caption{{ Risk profile with various $K_g$ of failing vehicles in a path graph.}}
    \label{fig:path_G}
\end{figure*}

\subsubsection{Failing Vehicles' Noise Magnitude}

In this research thrust, we analyze the effect of varying noise magnitudes, $g_i$, associated with failing vehicles $i \in \mathcal{I}_m$, which differ from the rest of the platoon. For non-failing vehicles, the noise magnitude is uniform, i.e., $g_j = g > 0$ for all $j \notin \mathcal{I}_m$, while for failing vehicles, the noise is scaled as $g_i = K_g \, g$ where $K_g > 0$ for all $i \in \mathcal{I}_m$. We explore several values of $K_g = \{0.01, 0.1, 1, 1.5, 2\}$, representing different noise levels experienced by the failing vehicles—either higher or lower compared to the rest of the platoon.

\vspace{1mm}

\noindent \underline{Complete Graph}: 
As shown in Fig. \ref{fig:complete_G}, in a complete graph, we set $g = 10$. It is observed that either an increased or decreased $K_g$ in the complete graph will negatively impact the vehicles that are not immediately next to the failure group. This is due to the fact that having a non-uniform term $K_g \, g_i$ in \eqref{eq:sigma_d} will break the balance among the statistics of the vehicles that are not immediate next to $\mathcal{I}_m$, since all non zero eigenvalues of the communication graph are exactly $n$. 

\vspace{1mm}
\noindent \underline{1-Cycle Graph}: 
In Fig. \ref{fig:cycle_G}, we set $g=1$ for a $1-$cycle graph. In an $1-$ cycle graph, changes of $K_g$ will not impact the risk of other vehicles as much as in the complete graph due to its low connectivity. However, similar to the complete graph, the growth of the failure group will exaggerate the impact since the contributions of the $\mathcal{I}_m$'s eigenvalue and eigenvectors are increasing.

\vspace{1mm}
\noindent \underline{Path Graph}:
In Fig. \ref{fig:path_G}, we set $g=0.1$ for a path graph. As the overall connectivity decreases, the path graph further restraints the impact from the failure group $\mathcal{I}_m$, it also presents a unique feature as the smaller failure group with $K_g < 1$ will negatively impact the other vehicles but the larger failure group with $K_g < 1$ will positively reduce the cascading risk of the other vehicles.

\subsubsection{Impacts of the Non-uniformity in Additive Noise}
\begin{figure*}[t]
    \centering
    \includegraphics[width=0.8\linewidth]{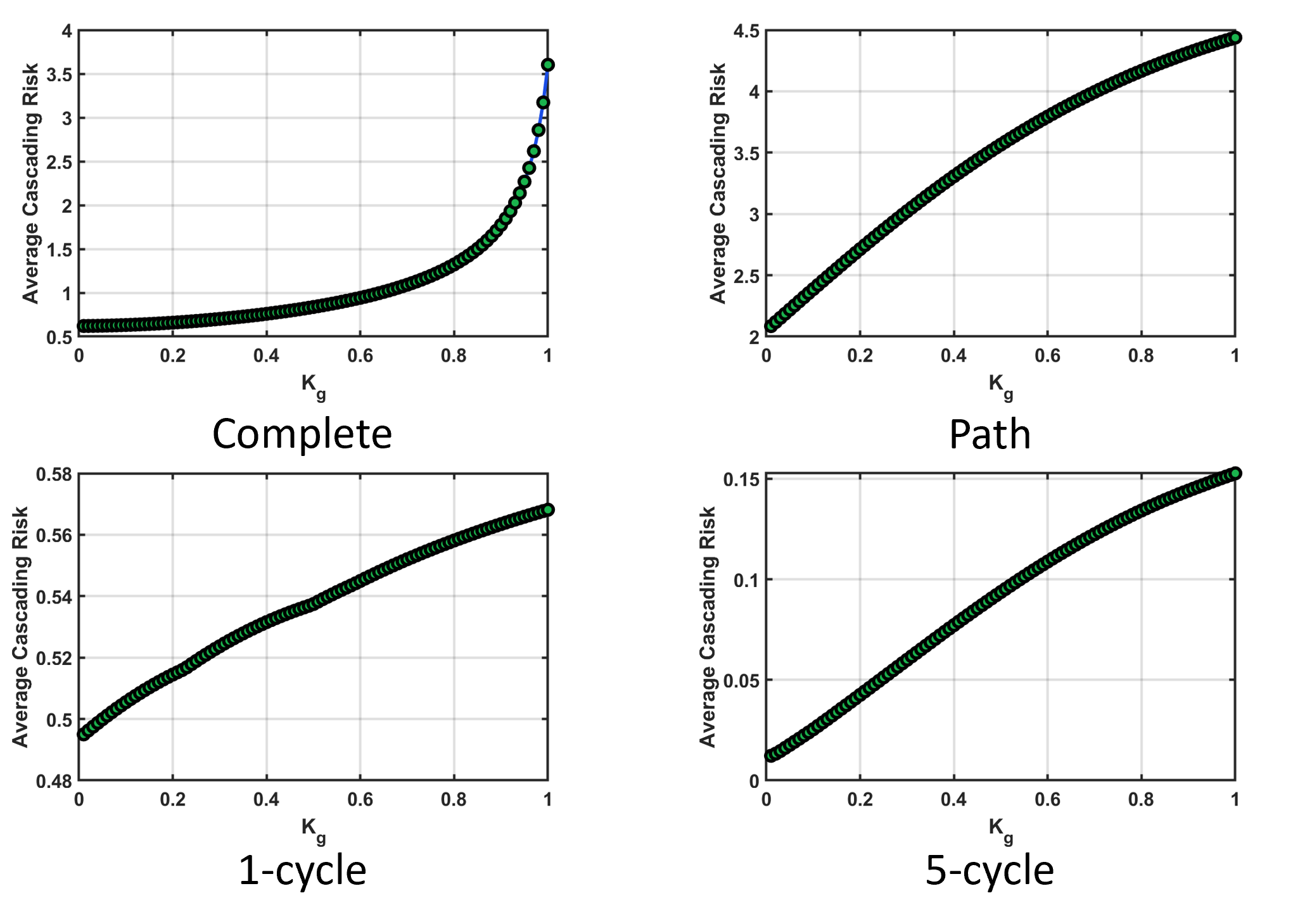}
    \caption{Average risk of cascading failures with various non-uniformity in the noise magnitudes.}
    \label{fig:G_sin}
\end{figure*}

In this research thrust, we analyze the effect of non-uniform noise magnitudes, $g_i$, across the platoon. For all vehicles, we define $g_i = (K_g \, \sin(i) + 1) \cdot g$ with $i = 1,...,n$, where $g > 0$, and $K_g \in [0.01,1]$ quantifies the overall non-uniformity in $g_i$. We explore several communication graphs and evaluate their resilience to variations in noise magnitudes.

In Fig. \ref{fig:G_sin}, we set $g = 10$ for the complete graph, $g = 1$ for the 5-cycle graph, $g = 0.1$ for the 1-cycle graph, and $g = 0.01$ for the path graph. When a collision occurs at the $10$'th vehicle pair, we observe that the complete graph shows the greatest resilience to increasing noise non-uniformity, with resilience decreasing as graph connectivity decreases, i.e., from cycle graphs to the path graph. This suggests that higher connectivity in the communication graph is more effective in mitigating cascading failures under non-uniform noise.

\section{Conclusion and Discussion}

For the time-delayed vehicle platooning problem defined in \cite{Somarakis2020b}, we introduce the notion of the risk of cascading collisions and extend it with the average value-at-risk measure ($\AVAR$) and with various scenarios on existing failures. We explore the properties of the cascading collision risk among several special communication graph topologies with symmetries by providing the closed-form representation of the risk of cascading collisions formula. Furthermore, we reveal and quantify the graph structure and time-delay induced limits on the risk of cascading collisions for general connected communication graphs and tighter limits on the complete graph. Our case study sheds light on the impact of existing failures, including their scale, distribution in the network, and the magnitude of the failure. From the communication graph design perspective, our result provides insight on the most advisable choice of mitigating the risk of cascading collisions when adding or removing links from the communication graphs.

The future directions of this work includes relaxing the assumption on the Gaussian input noise and investigating the distributionally robust risk of cascading collision in vehicle platooning \cite{liu2024data,pandey2023quantification}, the perception uncertainties \cite{liu2023symbolic,amini2023impact}, and extending the current framework from point mass multi-agents to multiple vehicle dynamics acting on the plane. One case study of interest is risk analysis in the dynamics of networked non-holonomic mobile vehicles for relevant formation control problems.

\appendix

\noindent \textbf{Algebraic Graph Theory:} A weighted graph is defined by $\mathcal{G} = (\mathcal{V}, \mathcal{E}, \omega)$, where $\mathcal{V}$ is the set of nodes, $\mathcal{E}$ is the set of edges (feedback links), and $\omega: \mathcal{V} \times \mathcal{V} \rightarrow \mathbb{R}_{+}$ is the weight function that assigns a non-negative number (feedback gain) to every link. Two nodes are directly connected if and only if $(i,j) \in \mathcal{E}$.
Every graph $\mathcal{G}$ is considered to be connected. In this paper, for every $i,j \in \mathcal{V}$, the following properties hold:
\begin{itemize}
    \item $\omega(i,j) > 0$ if and only if $(i,j) \in \mathcal{E}$.
    \item $\omega(i,j) = \omega(j,i)$, i.e., links are undirected.
    \item $\omega(i,i) = 0$, i.e., links are simple.
\end{itemize}

\noindent \textbf{Proof of Lemma \ref{lem:d_steady}: }
We refer to the proof of Theorem 1, \cite{liu2021risk}. \hfill$\square$
\vspace{0.1cm}

\noindent \textbf{Proof of Theorem \ref{thm:risk_levelset}: }Knowing that the steady-state distance of the $i$'th pair is observed to be $\bar{d}_i \in C_{\delta^*}$, we can compute the conditional probability of $\mathbb{P} \{\bar{d}_j < \hat{d} ~|~ \bar{d}_i < d^* \}$ as
\begin{equation*}
    \begin{aligned}
        \mathbb{P} \{\bar{d}_j < \hat{d} | \bar{d}_i < d^* \} = \frac{\mathbb{P} \{\bar{d}_j < \hat{d} \wedge \bar{d}_i < d^* \}}{\mathbb{P} \{\bar{d}_i < d^* \}}
        = \frac{\int_{-\infty}^{d^*} \int_{-\infty}^{\hat{d}} h(x,y) dxdy}{\frac{1}{2}\left(1+ \textrm{erf}(\frac{x-r}{\sqrt{2}\sigi}) \right)},
    \end{aligned}
\end{equation*}
for any $\hat{d} \in \R$, and $h(x,y)$ denotes the probability density function of a bi-variate normal distribution \cite{tong2012multivariate} such that
\begin{equation*}
        \begin{aligned}
            h(x, y) = \frac{\exp \left\{-\frac{1}{2(1-\rho^2_{ij})} \left[ \left(\frac{x-r}{\sigi} \right)^2 + \left(\frac{y-r}{\sigj} \right)^2 - 2\rho_{ij} \frac{(x-r)(y-r)}{\sigi \sigj}\right] \right\}}{2\pi \sigi \sigj \sqrt{1 -\rho^2_{ij}}},        
        \end{aligned}
\end{equation*}
where $\sigi, \sigj,$ and $\rho_{ij}$ are defined in Lemma \ref{lem:d_steady}. Then, the $\VAR$ is given by the solution of $\mathbb{P} \{\bar{d}_j < \hat{d} ~|~ \bar{d}_i < d^* \} = \varepsilon$ such that
\begin{equation*}
        \mathfrak{R}^{i,j}_\varepsilon = \text{sol} \left\{\hat{d} ~\Big|~ \int_{-\infty}^{d^*} \int_{-\infty}^{\hat{d}} h(x,y) ~ dxdy = \frac{\varepsilon}{2}\left(1+ \textrm{erf}(\frac{x-r}{\sqrt{2}\sigi}) \right) \right\}.
\end{equation*}
Following the definition of $\AVAR$, we have 
\begin{align*}
    \mathfrak{A}^{i,j}_\varepsilon = \frac{1}{\varepsilon} \frac{\int_{-\infty}^{d^*} \int_{-\infty}^{\mathfrak{R}^{i,j}_\varepsilon} y h(x,y) ~ dxdy}{\frac{1}{2}\left(1+ \textrm{erf}(\frac{x-r}{\sqrt{2}\sigi}) \right)}.
\end{align*}
Then, the $\cass$ can be computed by comparing the right boundary of $C_\delta$ with $\mathfrak{A}^{i,j}_\varepsilon$ which also quantifies all branches in Theorem \ref{thm:risk_levelset}.
\hfill$\square$

\vspace{0.1cm}

\noindent \textbf{Proof of Lemma \ref{lem:bivariate}:}
The result follows directly after Lemma \ref{lem:d_steady} and the conditional distribution of a bi-variate normal random variable as in \cite{tong2012multivariate}. \hfill$\square$

\vspace{0.1cm}

\noindent \textbf{Proof of Theorem \ref{thm:avar_single_cas}:} Given the fact that the continuous conditional random variable $\bar{d}_j|{\bar{d}_i = d^*}$ follows a normal distribution, $\Pro \{\bar{d}_j|{\bar{d}_i = d^*} < z\}$ can be computed as 
\[
    \frac{1}{2} \left(1+\textup{erf} \left(\frac{z-\tilde{\mu}}{\sqrt{2} \tilde{\sigma}} \right) \right),
\]
and the $\VAR$ is given by
\[
    \mathfrak{R}^{i,j}_\varepsilon = \sqrt{2} \tilde{\sigma} \textup{erf}^{-1}(2\varepsilon-1) + \tilde{\mu}.
\]
The corresponding $\AVAR$ is evaluated by using change of variable
\begin{align*}
    \mathfrak{A}^{i,j}_\varepsilon &= \frac{1}{\varepsilon} \int_{-\infty}^{\sqrt{2} \tilde{\sigma} \textup{erf}^{-1}(2\varepsilon-1) + \tilde{\mu}} z \frac{1}{\sqrt{2 \pi} \tilde{\sigma}} \exp \left( -\frac{(z-\tilde{\mu})^2}{2\tilde{\sigma}^2}\right)\textup{d}z\\
    & = \tilde{\mu} - \frac{\tilde{\sigma}}{\sqrt{2\pi} \varepsilon \exp(\io^2)},
\end{align*}
where $\io = \textrm{erf}^{-1} (2\varepsilon-1)$. Then, one can conclude the result by solving for $\mathfrak{A}_{\varepsilon}^{i,j} = \frac{r}{\delta+c}$ as well as comparing $\mathfrak{A}_{\varepsilon}^{i,j}$ with $0$ and $\frac{d}{c}$ to obtain branch conditions. \hfill$\square$

\vspace{0.1cm}

\noindent \textbf{Proof of Lemma \ref{lem:multi_condition}:}
The result follows directly after Lemma \ref{lem:d_steady} and the conditional distribution of a multi-variate normal random variable as in \cite{tong2012multivariate}. \hfill$\square$

\vspace{0.1cm}

\noindent \textbf{Proof of Corollary \ref{cor:risk_mul_cas}:} The result can be derived using the similar lines of argument as in Theorem \ref{thm:avar_single_cas} and apply the result from Lemma \ref{lem:multi_condition} for the conditional statistics. \hfill$\square$

\vspace{0.1cm}

\noindent \textbf{Proof of Lemma \ref{lem:sig_complete}:}
Following Lemma \ref{lem:d_steady}, given that the eigenvalues $\lambda_k$ are identical for all $k = 2,\dots,n$,
\begin{equation*}
    \begin{aligned}
        \sigma_{ij} &=  \frac{g^2\tau^3 f(n \tau, \beta \tau)}{2 \pi}\sum_{k=2}^{n} \big(\tilde{\bm{e}}_{i}^T \bm{q}_k \big) \big(\tilde{\bm{e}}_{j}^T \bm{q}_k \big)\\
        & = \frac{g^2\tau^3 f(n \tau, \beta \tau)}{2 \pi} \sum_{k=2}^{n} \phi(i,j,k),
    \end{aligned}
\end{equation*}
where 
\[
    \phi(i,j,k) = ( q_{k,i+1}q_{k,j+1} - q_{k,i+1}q_{k,j} - q_{k,i}q_{k,j+1} + q_{k,i}q_{k,j} ),
\]
and $q_{i,j}$ denotes the element of the matrix $Q = [\bm{q}_1 | \dots | \bm{q}_n]$. From the orthogonality property of the normalized eigenvectors, it is known that $\sum_{k} q_{k,i} q_{k,j} = \delta_{ij}$, in which $\delta_{ij}$ denotes the Kronecker delta. We have if $i = j$, then
\begin{align*}
    \sigma_{ij} = g^2 \frac{\tau^3}{2\pi} \, f(n \tau, \beta \tau) \big( 1 - 0 - 0 + 1\big) = \frac{1}{\pi} \, f(n \tau, \beta \tau) \, g^2 \, \tau^3.
\end{align*}
For $|i-j| = 1$, which is equivalent to $j = i+1$ or $i = j+1$,
$$
    \sigma_{ij} 
    = -\frac{1}{2\pi} \, f(n \tau, \beta \tau) \, g^2 \, \tau^3,
$$
 and for $|i-j| > 1$, one has
$
    \sigma_{ij} = 0.
$
\hfill$\square$

\vspace{0.1cm}

\noindent \textbf{Proof of Theorem \ref{thm:mul_risk_complete}:}
One can observe that the structure of $\tilde{\Sigma}_{22}$ is a tridiagonal matrix. Using the result from \cite{usmani1994inversion}, considering $\theta_{i} = 2^{-i} \sigma_c^i (i+1)$, the inversion $\tilde{\Sigma}_{22}^{-1} = [\alpha_{ij}]$ is given by:
\[
        \alpha_{ij}=\begin{cases}
            \left(\prod_{i}^{j-1} \frac{\sigma_c}{2}\right) \frac{\theta_{i-1}\theta_{m-j}}{\theta_{m}}, & i < j\\[5pt]
            \frac{2\, i \, (m+1-i)}{\sigma_c\, (m+1)}, & i = j\\[5pt]
            \left(\prod_{j+1}^{i} \frac{\sigma_c}{2}\right) \frac{\theta_{j-1}\theta_{m-i}}{\theta_{m}} , & i > j\\
            \end{cases}.
    \]

\noindent \underline{Case 1}: The result is immediate since $\tilde{\Sigma}_{12} = \tilde{\Sigma}_{21}^T = [0, \dots, 0]$, the conditional mean value and the variance remains unchanged.

\noindent \underline{Case 2}: Since the existing failures have no impact for the non-adjacent vehicle pairs, we only consider the adjacent failure group with size $m' \leq m$. Given that $\hat{\Sigma}_{12} = \hat{\Sigma}_{21}^T = [-\sigma_c/2,0,\dots,0]$ or $[0,\dots,0,-\sigma_c/2]$. Using the symmetry, the conditional variance can be written as
$$
    \hat{\sigma}^2 
    = \sigi^2 - \frac{\sigma^2_c}{4} \alpha_{11} = \sigi^2 - \frac{\sigma^2_c}{4} \alpha_{m'm'}
    = \sigi^2 - \frac{\sigma_c \, m'}{2(m'+1)},
$$
and the value of $\hat{\mu}$ can be obtained with the same lines of argument.

\noindent \underline{Case 3}: We consider the $j$'th pair is between two groups with the sizes as $m_1$ and $m_2$. Let us form a new $(m_1+m_2) \times (m_1+m_2)$ covariance matrix $\hat{\Sigma}$, and the rest of the result follows similarly as the previous case. 

Then, one can compute $\cass$ using the conditional statistics evaluated above and the similar lines of argument in the proof of Theorem \ref{thm:avar_single_cas}.
\hfill$\square$

\vspace{0.1cm}

\noindent \textbf{Proof of Theorem \ref{thm:sig_upper_lower_limit}:}
Let us rewrite \eqref{eq:sigma_d} as 
\begin{equation*}
    \begin{aligned}
        \sigma_{ij} &= g^2\frac{\tau^3}{2\pi} \Tr \big(\text{diag}\{\tilde{\bm{e}}_i^T\bm{q}_1\tilde{\bm{e}}_j^T\bm{q}_1 f(\lambda_1 \tau, \beta \tau),...,\\ 
        &\hspace{5cm} \tilde{\bm{e}}_i^T\bm{q}_n\tilde{\bm{e}}_j^T\bm{q}_nf(\lambda_n \tau, \beta \tau)\} \big)\\
        & = g^2\frac{\tau^3}{2\pi} \Tr \big(\text{diag}\left\{f(\lambda_1 \tau, \beta \tau),...,f(\lambda_n \tau, \beta \tau)\right\} \times \\
        & \hspace{3.5cm} \text{diag}\left\{ \tilde{\bm{e}}_i^T\bm{q}_1\tilde{\bm{e}}_j^T\bm{q}_1,...,\tilde{\bm{e}}_i^T\bm{q}_n\tilde{\bm{e}}_j^T\bm{q}_n \right\}\big)\\
        & = g^2\frac{\tau^3}{2\pi} \Tr ( F E_{ij} ) = g^2\frac{\tau^3}{2\pi} \Tr ( E_{ij}F ),
    \end{aligned}
\end{equation*}
where $F = \text{diag}\left\{\underline{f},f(\lambda_2 \tau, \beta \tau),...,f(\lambda_n \tau, \beta \tau)\right\}$, and $E_{ij} = (\tilde{\bm{e}}_i^T Q)^T \tilde{\bm{e}}_j^T Q$. Since $\tilde{\bm{e}}_i^T\bm{q}_1\tilde{\bm{e}}_j^T\bm{q}_1 = 0$ always holds, we can set $(F)_{11} = \underline{f}$, the lowerbound of $f$, without loss of generality. Considering the fact that $F$ is a normal matrix \cite{horn2012matrix} in $\R^{n \times n}$, we can use the result from \cite{liu2009trace} (Theorem 2.10) to show 
\[
    \sum_{i = 1}^{n} \re(\gamma_{n-i+1}(F))\mu_i(\bar{E}_{ij}) \leq \Tr(E_{ij}F),
\] 
and
\[
    \Tr(E_{ij}F) \leq \sum_{i = 1}^{n} \re(\gamma_{n-i+1}(F))\mu_{n-i+1}(\bar{E}_{ij}),
\]
where $\gamma_{i}(\cdot)$ and $\mu_{i}(\cdot)$ denotes the $i$'th eigenvalue of the matrix $F$ and $\bar{E}_{ij}$ in the non-decreasing order, $\bar{E}_{ij} = (E_{ij}+E_{ij}^T)/2$, and $\re(\cdot)$ denotes the real part of the eigenvalue. Let us denote by the eigenvalues of $F$ as $\gamma_i(F)$, the smallest and the largest eigenvalue as $\gamma_{min}(F)$ and $\gamma_{max}(F)$. Then, the above inequality can be written as
\begin{equation*}
    \begin{aligned}
        \sum_{i = 1}^{n} \gamma_i(F) \mu_i(\bar{E}_{ij}) \leq  \Tr(E_{ij}F) \leq \sum_{i = 1}^{n} \gamma_i(F) \mu_{n-i+1}(\bar{E}_{ij}).
    \end{aligned}
\end{equation*}
Considering the fact that
\[
    \bar{E}_{ij} = \frac{E_{ij}+E_{ij}^T}{2} = \frac{Q^T(\tilde{\bm{e}}_i\tilde{\bm{e}}_j^T+\tilde{\bm{e}}_j\tilde{\bm{e}}_i^T)Q}{2},
\]
one has $\mu_i(\bar{E}_{ij}) = \mu_i(QQ^T \frac{\tilde{\bm{e}}_i\tilde{\bm{e}}_j^T+\tilde{\bm{e}}_j\tilde{\bm{e}}_i^T}{2}) = \mu_i(\frac{\tilde{\bm{e}}_i\tilde{\bm{e}}_j^T+\tilde{\bm{e}}_j\tilde{\bm{e}}_i^T}{2})$. By observing the structure of $\tilde{E}_{ij} = \frac{\tilde{\bm{e}}_i\tilde{\bm{e}}_j^T+\tilde{\bm{e}}_j\tilde{\bm{e}}_i^T}{2}$, there exists a fixed number of rows and columns with zero entry based on the value of $|i-j|$. Finding the eigenvalue of $\tilde{E}_{ij}$ can be simplified as follows.

\noindent{\underline{Case 1}:} If $|i-j| = 0$, there exists $n-2$ rows and $n-2$ columns with zero entry such that
\begin{align*}
    \det (\mu I_n - \tilde{E}_{ij}) &= \mu^{n-2} \det \left(\mu I_2 - \begin{bmatrix}
    1 &-1 \\
    -1 &1 
    \end{bmatrix} \right) \\
    &= \mu^{n-1} (\mu - 2) = 0, 
\end{align*}
and $\mu_1 = 2$, $\mu_2 = ... = \mu_n = 0$.

\noindent{\underline{Case 2}:} If $|i-j| = 1$, there exists $n-3$ rows and $n-3$ columns with zero entry and 
\begin{align*}
    \det (\mu I_n - \tilde{E}_{ij}) &= \mu^{n-3} \det \left(\mu I_3 -  \begin{bmatrix}
    0 &1/2 &-1/2\\
    1/2 &1 &1/2\\
    -1/2 &1/2 &0
    \end{bmatrix} \right)\\
    & = \mu^{n-2} (\mu+\frac{3}{2})(\mu - \frac{1}{2}) = 0, 
\end{align*}
and $\mu_1 = 1/2$, $\mu_2 = ... = \mu_{n-1} = 0$, $\mu_n = -3/2$.

\noindent{\underline{Case 3}:} If $|i-j| > 1$, there exists $n-4$ rows and $n-4$ columns with zero entry and 
\begin{equation*}
    \begin{aligned}
        \det (\mu I_n - \tilde{E}_{ij})
        &= \mu^{n-4} \det \left(\mu I_4 -  \begin{bmatrix}
        0 & 0 &1/2 &-1/2\\
        0 & 0 &-1/2 &1/2\\
        1/2 &-1/2 &0 &0\\
        -1/2 &1/2 &0 &0
        \end{bmatrix} \right)\\
        & = \mu^{n-2} (\mu+1)(\mu - 1) = 0, 
    \end{aligned}
\end{equation*}
and $\mu_1 = \mu_n = 1$, $\mu_2 = ... = \mu_{n-1} = 0$. Then, by considering the fact that $\gamma_{min}(F) \geq \underline{f}$ and $\gamma_{max}(F) \leq \bar{f}$ for the { convex} and compact subset $\bar{S}$, one can conclude the result.
\hfill$\square$

\vspace{0.1cm}
\noindent \textbf{Proof of Theorem \ref{thm:cas_bound}:} 
Considering the fact that the risk of cascading collisions $\cass$ is a monotone function with respect to $\mathfrak{A}_\varepsilon$ when it takes nonnegative finite values, one can obtain the best achievable risk by evaluating the upper limit of $\mathfrak{A}_\varepsilon$. When $\bar{d}_i = 0$, using the result from Lemma \ref{lem:bivariate} and Theorem \ref{thm:avar_single_cas}, one can write $\mathfrak{A}_\varepsilon$ as a function of $\sigi,\sigj,\sigma_{ij}$ such that
\begin{equation*}
    \begin{aligned}
        \mathfrak{A}_\varepsilon (\sigi,\sigj,\sigma_{ij}) = r - r \, \frac{\sigma_{ij}}{\sigi^2} - \kappa_\varepsilon \, \sqrt{\sigj^2 - \frac{\sigma_{ij}^2}{\sigi^2}},
    \end{aligned}
\end{equation*}
where $\kappa_\varepsilon > 0$. Considering the fact that $\sigi,\sigj,\sigma_{ij}$, obtained from \eqref{eq:single_cond_stat}, are coupled through the graph structure, let us denote the set of all feasible combinations of the triplet $(\sigi,\sigj,\sigma_{ij})$ by $\mathbb{W}_{\mathcal{G}}$. The result from Theorem \ref{thm:sig_upper_lower_limit} also quantifies a feasible range for the triplet by 
\begin{align*}
    \mathbb{W} : =\Big\{ (\sigma_i,\sigma_j,\sigma_{ij})~:~ \sigma_i,\sigma_j \in \mathbb W_r \, , \sigma_{ij}\in \left(\max\{-\sigma_i\sigma_j,\underline{\sigma}_{\text{cov}}\},\min\{\sigma_i\sigma_j,\bar{\sigma}_{\text{cov}}\} \right) \Big\},
\end{align*}
with $\mathbb{W}_r = [\sqrt{2 \underline{\sigma}}, \sqrt{2 \bar{\sigma}}]$, $\underline{\sigma}_{\text{cov}}$ and $\bar{\sigma}_{\text{cov}}$ denotes the limits of $\sigma_{ij}$ depending on the values of $|i-j|$ as shown in Theorem \ref{thm:sig_upper_lower_limit}, respectively. One can show that $\mathbb{W}_{\mathcal{G}} \subset \mathbb{W}$ since $\mathbb{W}$ is not coupled through the graph structure. Knowing that $\mathfrak{A}_\varepsilon$ is continuous and bounded over $\mathbb{W}$, we have
\[
    \underset{(\sigi,\sigj,\sigma_{ij}) \in \mathbb{W}_{\mathcal{G}}}{\sup} \mathfrak{A}_\varepsilon \leq \underset{(\sigi,\sigj,\sigma_{ij}) \in \mathbb{W}}{\sup} \mathfrak{A}_\varepsilon.
\]
To facilitate the analysis, we will then consider the best achievable $\cass$ among $\mathbb{W}$. Knowing that the function $\mathfrak{A}_\varepsilon (\sigi,\sigj,\sigma_{ij})$ is continuous and differentiable functions of $\sigi>0$, $\sigj >0$, and $\sigma_{ij} \in (-\sigi \sigj,\sigi \sigj)$, the gradient of $\mathfrak{A}_\varepsilon$ is computed as
\[
    \nabla(\mathfrak{A}_\varepsilon) = \left[\frac{\partial \mathfrak{A}_\varepsilon}{\partial \sigi}, \frac{\partial \mathfrak{A}_\varepsilon}{\partial \sigj}, \frac{\partial \mathfrak{A}_\varepsilon}{\partial \sigma_{ij}} \right]^T.
\]

Observing that the term $\frac{\partial \mathfrak{A}_\varepsilon}{\partial \sigj} = -\frac{\kappa_\epsilon \sigj}{\sqrt{\sigj^2-\sigma_{ij}^2 / \sigi^2}}$ is strictly negative for all $\sigj \in \mathbb{W}_r$, the first-order optimality condition \cite{nocedal1999numerical} will not be satisfied within $\mathbb{W}$, indicating that the extreme values will only be obtained on the boundary of $\mathbb{W}$, which is given by
\begin{align*}
    \Big\{ (\sigi,\sigj,\sigma_{ij}) : \{\sigi = \sqrt{2 \underline{\sigma}}\text{ or } \sqrt{2 \bar{\sigma}}\} \cup \{\sigj = \sqrt{2 \underline{\sigma}} \text{ or }\sqrt{2 \bar{\sigma}} \}
    \\
    \cup \big\{\sigma_{ij} = \max\{-\sigma_i\sigma_j,\underline{\sigma}_{\text{cov}}\} \text{ or } \min\{\sigma_i\sigma_j,\bar{\sigma}_{\text{cov}}\} \big\} \Big\} \bigcap \mathbb{W}.
\end{align*}

Considering the fact that
\[  
    \frac{\partial^2 \mathfrak{A}_\varepsilon}{\partial \sigma_{ij}^2} = \frac{\kappa_\varepsilon}{\sigi ^2 (\sigj^2- \sigma_{ij}^2 / \sigi^2)^\frac{3}{2}} > 0,
\]
one can show that when $\sigma_{ij} > 0$, on the surface $\mathbb{W} \bigcap \{\sigi = \sqrt{2 \underline{\sigma}}\} \in \R^3$, the following inequality will hold
\[
    \mathfrak{A}_\varepsilon(\sqrt{2 \underline{\sigma}}, \sigj,\sigma_{ij}) \leq \mathfrak{A}_\varepsilon(\sqrt{2 \underline{\sigma}}, \sigj,\min\{\sigma_i\sigma_j,\bar{\sigma}_{\text{cov}}\}).
\]

\noindent Then, by repeating these steps for $\sigi = \sqrt{2 \bar{\sigma}}$, $\sigj = \sqrt{2 \underline{\sigma}}$, $ \sigj = \sqrt{2 \bar{\sigma}}$, and different signs of $\sigma_{ij}$, we can show that the extreme values of $\mathfrak{A}_\varepsilon$ lands on either surface $(\sigi,\sigj, \min\{\sigma_i\sigma_j,\bar{\sigma}_{\text{cov}}\})$ or surface $(\sigi,\sigj, \max\{-\sigma_i\sigma_j,\underline{\sigma}_{\text{cov}}\})$.

In the case when $\sigma_{ij} > 0$, one has $\frac{\partial \mathfrak{A}_\varepsilon}{\partial \sigj}|_{\sigma_{ij} = \sigi \sigj} < 0$ and $\frac{\partial \mathfrak{A}_\varepsilon}{\partial \sigj}|_{\sigma_{ij} = \bar{\sigma}_{\text{cov}}} < 0$, implying that the upper limit of $\mathfrak{A}_\varepsilon$ can only be achieved when $\sigj = \sqrt{2 \underline{\sigma}}$, such that 
\begin{equation*}
    \begin{aligned}
        \underset{(\sigi,\sigj,\sigma_{ij}) \in \mathbb{W}, \sigma_{ij} > 0}{\sup} \mathfrak{A}_\varepsilon &< \underset{\sigi \in [\sqrt{2 \underline{\sigma}}, \sqrt{2 \bar{\sigma}}]}{\sup} \mathfrak{A}_\varepsilon(\sigi, \sqrt{2 \underline{\sigma}}, \sigi \sqrt{2 \underline{\sigma}})\\
        &= \mathfrak{A}_\varepsilon(\sqrt{2 \bar{\sigma}}, \sqrt{2 \underline{\sigma}}, 2\sqrt{\bar{\sigma}\underline{\sigma}})\\
        &=  r\left(1-\sqrt{\underline{\sigma}} / \sqrt{\bar{\sigma}} \right).
    \end{aligned}
\end{equation*}

In the case when $\sigma_{ij} < 0$, one has $\frac{\partial \mathfrak{A}_\varepsilon}{\partial \sigj}|_{\sigma_{ij} = \sigi \sigj} > 0$ and $\frac{\partial \mathfrak{A}_\varepsilon}{\partial \sigj}|_{\sigma_{ij} = \underline{\sigma}_{\text{cov}}} < 0$. Considering the fact that $0>\underline{\sigma}_{\text{cov}} > -2\bar{\sigma}$ despite the value of $|i-j|$, one can show that the upper limit of $\mathfrak{A}_\varepsilon$ will be achieved when $\sigma_{ij} = \underline{\sigma}_{\text{cov}} = -\sigi \sigj$, such that 
\begin{equation*}
    \begin{aligned}
        \underset{(\sigi,\sigj,\sigma_{ij}) \in \mathbb{W}, \sigma_{ij} < 0}{\sup} \mathfrak{A}_\varepsilon &<  \underset{\sigi \in [\sqrt{2 \underline{\sigma}}, \sqrt{2 \bar{\sigma}}]}{\sup} \mathfrak{A}_\varepsilon \left(\sigi, \frac{|\underline{\sigma}_{\text{cov}}|}{\sigi}, \underline{\sigma}_{\text{cov}} \right)\\
        &= \mathfrak{A}_\varepsilon \left(\sqrt{2 \underline{\sigma}}, \frac{|\underline{\sigma}_{\text{cov}}|}{\sqrt{2 \underline{\sigma}}}, \underline{\sigma}_{\text{cov}} \right)\\
        &=  r\left(1-\underline{\sigma}_{\text{cov}} / 2\underline{\sigma} \right).
    \end{aligned}
\end{equation*}
Considering the fact that $\underline{\sigma}_{\text{cov}} < 0$, the upper limit of $\mathfrak{A}_\varepsilon$ when $\sigma_{ij} < 0$ will be strictly greater than $r$, indicating the best achievable $\cass$ is always $0$ in this case.

In addition to the above scenarios, in the special case when $\sigma_{ij} = 0$, the best achievable risk of cascading collisions boils down to the best achievable risk of single collision as introduced in \cite{Somarakis2020b}, which solely depends on the value of $\sqrt{2 \underline{\sigma}}$. Then, one can conclude by converting the limit of  $\mathfrak{A}_\varepsilon(\sigi,\sigj,\sigma_{ij})$ into the limit of $\cass$. 
\hfill$\square$

\vspace{0.1cm}
    
\noindent \textbf{Proof of Corollary \ref{cor:limit_complete}:} Observing the structure of the steady-state covariance terms (Lemma \ref{lem:sig_complete}), the risk of cascading collision $\cass$ can be considered in two cases based on the value of $|i-j|$. In the case when $|i-j| > 1$, $\cass$ boils down to the risk of single collision, and the best achievable risk can be obtained at $\sigj = \sqrt{2\underline{\sigma}}$. In the case when $|i-j| = 1$, $\mathfrak{A}_\varepsilon$ can be written as
\[
    \mathfrak{A}_\varepsilon = \frac{3}{2} r - \frac{1}{2} d^* - \kappa_\varepsilon \sqrt{\frac{3}{4} \sigma_c},
\]
where $\sigma_c$ is shown in Lemma \ref{lem:sig_complete}. The upper limit of $\mathfrak{A}_\varepsilon$ can be obtained as $\sigma_c \rightarrow 2\underline{\sigma}$. Then, one can conclude by converting the limit of $\mathfrak{A}_\varepsilon(\sigi,\sigj,\sigma_{ij})$ into the limit of $\cass$.
\hfill$\square$


\printbibliography

@book{Evans2013,
    title = {{An Introduction to Stochastic Differential Equations}},
    year = {2013},
    booktitle = {An Introduction to Stochastic Differential Equations},
    author = {Evans, Lawrence},
    month = {12},
    publisher = {American Mathematical Society}
}

@book{van2010graph,
    title = {{Graph spectra for complex networks}},
    year = {2010},
    author = {Van Mieghem, Piet},
    publisher = {Cambridge University Press}
}

@book{Follmer2016,
    title = {{Stochastic Finance}},
    year = {2016},
    booktitle = {Stochastic Finance},
    author = {F{\"{o}}llmer, Hans and Schied, Alexander},
    month = {7},
    publisher = {De Gruyter}
}

@book{mohammed1984stochastic,
    title = {{Stochastic functional differential equations}},
    year = {1984},
    author = {Mohammed, Salah-Eldin A and Salah-El Din, A Mohammed},
    volume = {99},
    publisher = {Pitman Advanced Publishing Program}
}

@book{tong2012multivariate,
    title = {{The multivariate normal distribution}},
    year = {2012},
    author = {Tong, Yung Liang},
    publisher = {Springer Science {\&} Business Media}
}

@article{gray2006toeplitz,
    title = {{Toeplitz and circulant matrices: A review}},
    year = {2006},
    author = {Gray, Robert M},
    publisher = {now publishers inc}
}

@inproceedings{ali2015enhanced,
  title={Enhanced flatbed tow truck model for stable and safe platooning in the presences of lags, communication and sensing delays},
  author={Ali, A. and Garcia, G. and Martinet, P.},
  booktitle={2015 IEEE International Conference on Robotics and Automation (ICRA)},
%   pages={1648--1653},
  year={2015},
  organization={IEEE}
}

@inproceedings{verginis2015decentralized,
  title={Decentralized 2-D control of vehicular platoons under limited visual feedback},
  author={Verginis, C. K. and Bechlioulis, C. P. and Dimarogonas, D. V. and Kyriakopoulos, K. J.},
  booktitle={2015 IEEE/RSJ International Conference on Intelligent Robots and Systems (IROS)},
%   pages={3566--3571},
  year={2015},
  organization={IEEE}
}

@inproceedings{tan1998demonstration,
  title={Demonstration of an automated highway platoon system},
  author={Tan, H. and Rajamani, R. and Zhang, W.},
  booktitle={Proceedings of the 1998 American control conference.},
  volume={3},
%   pages={1823--1827},
  year={1998},
  organization={IEEE}
}

@article{wu2017flow,
  title={Flow: Architecture and benchmarking for reinforcement learning in traffic control},
  author={Wu, C. and Kreidieh, Aboudy and Parvate, Kanaad and Vinitsky, Eugene and Bayen, Alexandre M},
  journal={arXiv preprint arXiv:1710.05465},
  volume={10},
  year={2017}
}

@article{bamieh2012coherence,
  title={Coherence in large-scale networks: Dimension-dependent limitations of local feedback},
  author={Bamieh, B. and Jovanovic, Mihailo R and Mitra, Partha and Patterson, Stacy},
  journal={IEEE Transactions on Automatic Control},
  volume={57},
  number={9},
  pages={2235--2249},
  year={2012},
  publisher={IEEE}
}

@article{dorfler2012synchronization,
  title={Synchronization and transient stability in power networks and nonuniform Kuramoto oscillators},
  author={Dorfler, F. and Bullo, F.},
  journal={SIAM Journal on Control and Optimization},
  volume={50},
  number={3},
  pages={1616--1642},
  year={2012},
  publisher={SIAM}
}

@article{acemoglu2015systemic,
  title={Systemic risk and stability in financial networks},
  author={Acemoglu, D. and Ozdaglar, A. and Tahbaz-Salehi, A.},
  journal={American Economic Review},
  volume={105},
  number={2},
  pages={564--608},
  year={2015}
}

@article{yu2010some,
  title={Some necessary and sufficient conditions for second-order consensus in multi-agent dynamical systems},
  author={Yu, W. and Chen, G. and Cao, M.},
  journal={Automatica},
  volume={46},
  number={6},
  pages={1089--1095},
  year={2010},
  publisher={Elsevier}
}

@ARTICLE{7438924,
  author={M. {Rahnamay-Naeini} and M. M. {Hayat}},
  journal={IEEE Transactions on Smart Grid}, 
  title={Cascading Failures in Interdependent Infrastructures: An Interdependent Markov-Chain Approach}, 
  year={2016},
  volume={7},
  number={4},
  pages={1997-2006},
  }

@article{zhang2019robustness,
  title={Robustness of interdependent cyber-physical systems against cascading failures},
  author={Zhang, Y. and Ya{\u{g}}an, O.},
  journal={IEEE Transactions on Automatic Control},
  volume={65},
  number={2},
  pages={711--726},
  year={2019},
  publisher={IEEE}
}

@article{zhang2018cascading,
  title={Cascading failures in interdependent systems under a flow redistribution model},
  author={Zhang, Y. and Arenas, A. and Ya{\u{g}}an, O.},
  journal={Physical Review E},
  volume={97},
  number={2},
  pages={022307},
  year={2018},
  publisher={APS}
}

@inproceedings{grunberg2017determining,
  title={Determining collision potential as a measure of robustness in vehicular networks},
  author={Grunberg, T. W and Gayme, D. F},
  booktitle={2017 American Control Conference (ACC)},
  pages={3992--3998},
  year={2017},
  organization={IEEE}
}

@article{siami2017growing,
  title={Growing linear dynamical networks endowed by spectral systemic performance measures},
  author={Siami, M. and Motee, N.},
  journal={IEEE Transactions on Automatic Control},
  volume={63},
  number={7},
  pages={2091--2106},
  year={2017},
  publisher={IEEE}
}

@inproceedings{somarakis2018risk,
  title={Risk of collision in a vehicle platoon in presence of communication time delay and exogenous stochastic disturbance},
  author={Somarakis, C. and Ghaedsharaf, Y. and Motee, N.},
  booktitle={2018 IEEE Conference on Decision and Control (CDC)},
  pages={4487--4492},
  year={2018},
  organization={IEEE}
}

@article{rockafellar2002conditional,
    title = {{Conditional value-at-risk for general loss distributions}},
    year = {2002},
    journal = {Journal of Banking and Finance},
    author = {Rockafellar, R. Tyrrell and Uryasev, Stanislav},
    number = {7},
    pages = {1443--1471},
    volume = {26}
}

@article{rockafellar2000optimization,
    title = {{Optimization of Conditional Value-at-Risk}},
    year = {1999},
    journal = {Portfolio The Magazine Of The Fine Arts},
    author = {Rockafellar, R. T. and Uryasev, S.},
    pages = {1--26},
    volume = {2}
}

@article{Somarakis2020b,
    title = {{Risk of Collision and Detachment in Vehicle Platooning: Time-Delay-Induced Limitations and Tradeoffs}},
    year = {2020},
    journal = {IEEE Transactions on Automatic Control},
    author = {Somarakis, C. and Ghaedsharaf, Y. and Motee, N.},
    number = {8},
    volume = {65}
}

@article{Somarakis2019g,
    title = {{Time-delay origins of fundamental tradeoffs between risk of large fluctuations and network connectivity}},
    year = {2019},
    journal = {IEEE Transactions on Automatic Control},
    author = {Somarakis, C. and Ghaedsharaf, Y. and Motee, N.},
    number = {9},
    volume = {64}
}

@inproceedings{liu2021risk,
  title={Risk of Cascading Failures in Time-Delayed Vehicle Platooning},
  author={Liu, Guangyi and Somarakis, Christoforos and Motee, Nader},
  booktitle={2021 60th IEEE Conference on Decision and Control (CDC)},
  pages={4841--4846},
  year={2021},
  organization={IEEE}
}

@article{usmani1994inversion,
  title={Inversion of a tridiagonal Jacobi matrix},
  author={Usmani, R. A.},
  journal={Linear Algebra and its Applications},
  volume={212},
  number={213},
  pages={413--414},
  year={1994},
  publisher={New York: Elsevier Science, 1968-}
}

@article{siami2016fundamental,
  title={Fundamental limits and tradeoffs on disturbance propagation in linear dynamical networks},
  author={Siami, Milad and Motee, Nader},
  journal={IEEE Transactions on Automatic Control},
  volume={61},
  number={12},
  pages={4055--4062},
  year={2016},
  publisher={IEEE}
}

@inproceedings{liu2022emergence,
  title={Emergence of Cascading Risk and Role of Spatial Locations of Collisions in Time-Delayed Platoon of Vehicles},
  author={Liu, Guangyi and Somarakis, Christoforos and Motee, Nader},
  booktitle={2022 IEEE 61st Conference on Decision and Control (CDC)},
  pages={6460--6465},
  year={2022},
  organization={IEEE}
}

@INPROCEEDINGS{liu2022risk,
  author={Liu, Guangyi and Pandey, Vivek and Somarakis, Christoforos and Motee, Nader},
  booktitle={2022 American Control Conference (ACC)}, 
  title={Risk of Cascading Failures in Multi-agent Rendezvous with Communication Time Delay}, 
  year={2022},
  volume={},
  number={},
  pages={2172-2177}}

@incollection{sarykalin2008value,
  title={Value-at-risk vs. conditional value-at-risk in risk management and optimization},
  author={Sarykalin, Sergey and Serraino, Gaia and Uryasev, Stan},
  booktitle={State-of-the-art decision-making tools in the information-intensive age},
  pages={270--294},
  year={2008},
  publisher={Informs}
}

@inproceedings{liu2023cascading,
  title={Cascading Waves of Fluctuation in Time-delay Multi-agent Rendezvous},
  author={Liu, Guangyi and Pandey, Vivek and Somarakis, Christoforos and Motee, Nader},
  booktitle={2023 American Control Conference (ACC)},
  pages={4110--4115},
  year={2023},
  organization={IEEE}
}

@book{horn2012matrix,
  title={Matrix analysis},
  author={Horn, Roger A and Johnson, Charles R},
  year={2012},
  publisher={Cambridge university press}
}

@article{liu2009trace,
  title={Trace inequalities for matrix products and trace bounds for the solution of the algebraic Riccati equations},
  author={Liu, Jianzhou and Zhang, Juan and Liu, Yu},
  journal={Journal of Inequalities and Applications},
  volume={2009},
  pages={1--17},
  year={2009},
  publisher={Springer}
}

@book{gu2003stability,
  title={Stability of time-delay systems},
  author={Gu, Keqin and Chen, Jie and Kharitonov, Vladimir L},
  year={2003},
  publisher={Springer Science \& Business Media}
}

@misc{zhou1996robust,
  title={Robust and optimal control},
  author={Zhou, Kemin and Doyle, John C and Glover, Keith},
  year={1996},
  publisher={Prentice-Hall, Inc.}
}

@article{sandberg2015cyberphysical,
  title={Cyberphysical security in networked control systems: An introduction to the issue},
  author={Sandberg, Henrik and Amin, Saurabh and Johansson, Karl Henrik},
  journal={IEEE Control Systems Magazine},
  volume={35},
  number={1},
  pages={20--23},
  year={2015},
  publisher={IEEE}
}

@article{sandberg2022secure,
  title={Secure networked control systems},
  author={Sandberg, Henrik and Gupta, Vijay and Johansson, Karl H},
  journal={Annual Review of Control, Robotics, and Autonomous Systems},
  volume={5},
  pages={445--464},
  year={2022},
  publisher={Annual Reviews}
}

@inproceedings{liu2024data,
  title={Data-Driven Distributionally Robust Mitigation of Risk of Cascading Failures},
  author={Liu, Guangyi and Amini, Arash and Pandey, Vivek and Motee, Nader},
  booktitle={2024 American Control Conference (ACC)},
  pages={3264--3269},
  year={2024},
  organization={IEEE}
}

@inproceedings{pandey2023quantification,
  title={Quantification of Distributionally Robust Risk of Cascade of Failures in Platoon of Vehicles},
  author={Pandey, Vivek and Liu, Guangyi and Amini, Arash and Motee, Nader},
  booktitle={2023 62nd IEEE Conference on Decision and Control (CDC)},
  pages={7401--7406},
  year={2023},
  organization={IEEE}
}

@book{nocedal1999numerical,
  title={Numerical optimization},
  author={Nocedal, Jorge and Wright, Stephen J},
  year={1999},
  publisher={Springer}
}

@inproceedings{ghaedsharaf2018performance,
  title={Performance of second-order platoon of vehicles in presence of time-delay and noise},
  author={Ghaedsharaf, Yaser and Somarakis, Christoforos and Motee, Nader},
  booktitle={2018 Annual American Control Conference (ACC)},
  pages={4887--4892},
  year={2018},
  organization={IEEE}
}

@article{strecok1968calculation,
  title={On the calculation of the inverse of the error function},
  author={Strecok, Anthony},
  journal={Mathematics of Computation},
  volume={22},
  number={101},
  pages={144--158},
  year={1968}
}

@inproceedings{amini2023impact,
  title={Impact of Misperception on Emergence of Risk in Platoon of Autonomous Vehicles},
  author={Amini, Arash and Liu, Guangyi and Pandey, Vivek and Motee, Nader},
  booktitle={2023 62nd IEEE Conference on Decision and Control (CDC)},
  pages={6086--6091},
  year={2023},
  organization={IEEE}
}

@inproceedings{liu2023symbolic,
  title={Symbolic perception risk in autonomous driving},
  author={Liu, Guangyi and Kamale, Disha and Vasile, Cristian-Ioan and Motee, Nader},
  booktitle={2023 American Control Conference (ACC)},
  pages={4077--4082},
  year={2023},
  organization={IEEE}
}

@article{amini2022space,
  title={Space--Time Sampling for Network Observability},
  author={Amini, Arash and Mousavi, Hossein K and Sun, Qiyu and Motee, Nader},
  journal={IEEE Transactions on Control of Network Systems},
  volume={10},
  number={3},
  pages={1159--1171},
  year={2022},
  publisher={IEEE}
}

\end{document}